\documentclass[proceedings]{JHEP3}
\usepackage{amssymb}
\usepackage{amsfonts}
\usepackage{amsmath}
\usepackage{epsfig}
\usepackage{xcolor}

\newbox\mybox

\newcommand\fverb{\setbox\mybox=\hbox\bgroup\verb}
\newcommand\fverbdo{\egroup\medskip\noindent\fbox{\unhbox\mybox}\ }
\newcommand\fverbit{\egroup\item[\fbox{\unhbox\mybox}]}
\conference{N-Extended Lorentzian Kac-Moody algebras}
\abstract{We investigate a class of Kac-Moody algebras previously not considered. We refer to them as n-extended Lorentzian Kac-Moody algebras
defined by their Dynkin diagrams through the connection of an $A_n$ Dynkin diagram to the node corresponding to the affine root. The cases $n=1$ and $n=2$
correspond to the well studied over and very extended Kac-Moody algebras, respectively, of which the particular examples of $E_{10}$ and $E_{11}$ play
a prominent role in string and M-theory. We construct closed generic expressions for their associated roots, fundamental weights and Weyl vectors. We use 
these quantities to calculate specific constants from which the nodes can be determined that when deleted decompose the n-extended Lorentzian Kac-Moody algebras
into simple Lie algebras and Lorentzian Kac-Moody algebra. The signature of these constants also serves to establish whether the algebras possess $SO(1,2)$ 
and/or $SO(3)$-principal subalgebras.}

\title{N-Extended Lorentzian Kac-Moody algebras}
\author{Andreas Fring and Samuel Whittington \\
Department of Mathematics, City, University of London,\\
Northampton Square, London EC1V 0HB, UK\\
E-mail: a.fring@city.ac.uk,samuel.whittington@city.ac.uk}

\input{tcilatex}
\begin{document}

\section{Introduction}

The symmetry algebras relevant in the formulation of fundamental theories in
particle physics have become increasingly complex over the years. While
finite dimensional Lie algebras are sufficient for the characterization of
local gauge symmetries describing three of the four known fundamental forces
in nature, it was noticed well over thirty years ago that infinite
dimensional Kac-Moody algebras \cite{kacinfinite} are needed for an adequate
description in the context of some string and conformal field theories \cite%
{goddard1985algebras,goddard1986kac,fring2018lectures}. For some string
theories, type II superstring theories or M-theory \cite{witten1995string},
a particular type of Lorentzian Kac-Moody algebras has turned out to be
especially relevant \cite%
{west2000hidden,west2001e11,damour200210,riccioni2007e11,englert2003s,bergshoeff2007e11,de2001hidden,berman2011,de2008gauged,hohm2014}%
.

In general, from a mathematical point of view the understanding of Kac-Moody
algebras is still partially incomplete \cite{kacinfinite}. However, based on
their Dynkin diagrams, that encode the structure of their corresponding
Cartan matrices, see e.g. \cite{Hum}, many subclasses have been fully
classified.\ The best known and studied subclasses are semisimple Lie
algebras of finite or affine type characterized by finite connected Dynkin
diagrams. Their Cartan matrices are positive definite in the former and
positive semi-definite in the latter case. In addition, hyperbolic Kac-Moody
algebras have also been fully classified \cite{carbone2010}. In terms of
their connected Dynkin diagrams they are defined by the property that the
deletion of \emph{any one node} leaves a possibly disconnected set of
connected Dynkin diagrams each of which is of finite type, except for at
most one affine type. Their Cartan matrices are nonsingular with exactly one
negative eigenvalue, i.e. they are Lorentzian.

While $E_{10}$ is a hyperbolic Kac-Moody algebra \cite{gebert199510}, $%
E_{11} $ is not \cite{nicolai2004low}, which, partially motivated by string
theory, led to the study of a larger class of Kac-Moody algebras that are
also Lorentzian \cite{borcherds1988generalized,gaberdiel2002class}. In \cite%
{gaberdiel2002class} these algebras were characterized in terms of their
connected Dynkin diagrams such that the deletion of \emph{at least one node}
leaves a possibly disconnected set of connected Dynkin diagrams each of
which is of finite type, except for at most one affine type. This definition
is obviously more general than the one for hyperbolic Kac-Moody algebras,
including them as subcases.

Of these algebras a particular type is very distinct. Referring to Dynkin
diagrams of the affine algebras as \emph{extended}, the \emph{over extended}
Dynkin diagrams consist of connecting a node to the affine root and the 
\emph{very extended} ones of connecting another new node to this new root.
Here we study root lattices resulting from Dynkin diagrams for which these
extensions are continued by adding successively nodes to the previous ones
and refer to them as $n$\emph{-extended} Dynkin diagrams. Our notation is
such that $n=0$ corresponds to the extended system,\ $n=1$ to the over
extended system, $n=2$ to the very extended system and $n>2$ to new systems
previously not studied. As we shall see below these algebras occur naturally
in the decomposition of the over and very extended systems, for instance the
decomposition of the over extended algebra $D_{17}$ contains a $5$-extended $%
E_{8}$-algebra.

When decomposing the $n$-extended algebras we encounter reduced Dynkin
diagrams which consist of an $A_{r}$-Dynkin diagram with an $A_{n}$-Dynkin
diagram attached to the $m$-th node. We denote the corresponding algebras by 
$\hat{A}_{r}^{(n,m)}$. We study the corresponding weight lattices in more
detail with a special focus on the case in which an $A_{n}$-Dynkin is
attached to the middle node of $A_{r}$ referring to it as $\hat{A}_{r}^{(n)}$%
.

Our manuscript is organised as follows: In section 2 we recall some known
facts about the extended, over extended and very extended root lattices to
establish our notations and conventions. In section 3 we define the new $n$%
-extended Lorentzian algebras and construct their roots, fundamental weights
and Weyl vectors. In section 4 we discuss the necessary criteria for the
occurrence of $SO(1,2)$ and $SO(3)$ principal subalgebras. In section 5 we
compute a special set of constants obtained from the inner product of the
Weyl vector and a fundamental weight, whose overall signs provide necessary
and sufficient conditions for the occurrence of $SO(1,2)$ and $SO(3)$
principal subalgebras and the decomposition of the $n$-extended Lorentzian
algebras, which are then studied in detail in section 6. Section 7 contains
a similar analysis to the one in sections 5 and 6 for the $\hat{A}%
_{r}^{(n,m)}$-algebras. Our conclusions are stated in section 8.

\section{Preliminaries, extended, over extended and very extended root
lattices}

Before we present our extended version of the Lorentzian Kac-Moody algebra
we recall some of the known results on the extended, over extended and very
extended root lattices to establish our conventions and notations. There
exist various types of choices to define the corresponding root spaces,
especially with regard to the selection of the inner product in the
corresponding vector space \cite{brown2004m,henneaux2008}. Here we adopt
most of the conventions used and introduced in \cite%
{nicolai2001dio,gaberdiel2002class,goddard1985algebras}.

The root lattice $\Lambda $ for a Lorentzian Kac-Moody algebra $\mathbf{g}$
consists of two parts. The first, $\Lambda _{\mathbf{g}}$, is spanned by the
simple roots $\alpha _{i}$, $i=1,\ldots ,r$, of the semisimple Lie algebra $%
\mathbf{g}$ with rank $r$. The second is the self-dual Lorentzian lattice $%
\Pi ^{1,1}$ equipped with the inner product%
\begin{equation}
z\cdot w=-z^{+}w^{-}-z^{-}w^{+}  \label{ip}
\end{equation}
for $z,w\in \Pi ^{1,1}$ of the form $z=(z^{+},z^{-})$, $w=(w^{+},w^{-})$.
There are two primitive null vectors in $\Pi ^{1,1}$, that will be important
below, $k=(1,0)$, $\bar{k}=(0,-1)$, with $k\cdot k=\bar{k}\cdot \bar{k}=0$, $%
k\cdot \bar{k}=1$ and two vectors $\pm \left( k+\bar{k}\right) $ of length $%
2 $. An \emph{extended}, or affine, root lattice is obtained by adding to
the set of simple roots the negative of the highest root $\theta
:=\sum\nolimits_{i=1}^{r}n_{i}\alpha _{i}$, with \emph{Kac labels} $n_{i}\in 
\mathbb{N}$. Here we add the modified negative highest root $\alpha
_{0}=k-\theta $ to obtain a differently extended root lattice $\Lambda _{%
\mathbf{g}_{0}}=\Lambda _{\mathbf{g}}\oplus \Pi ^{1,1}$. Adding to this set
of roots the root $\alpha _{-1}=-\left( k+\bar{k}\right) $ we obtain the 
\emph{over extended} root lattice $\Lambda _{\mathbf{g}_{-1}}=\Lambda _{%
\mathbf{g}_{0}}\oplus \Pi ^{1,1}$ and adding the root $\alpha _{-2}=k-\left(
\ell +\bar{\ell}\right) $ produces the \emph{very extended} root lattice $%
\Lambda _{\mathbf{g}_{-2}}=\Lambda _{\mathbf{g}_{-1}}\oplus \Pi ^{1,1}$.
Here $\ell $, $\bar{\ell}$ are two primitive null vectors in the second
self-dual Lorentzian lattice.

We summarize these properties in the following table.

\begin{center}
\begin{tabular}{c|l|l|l|l}
algebra & root lattice & added root & Dynkin diagram & expl. \\ \hline
$\mathbf{g}_{0}$ & $\Lambda _{\mathbf{g}_{0}}=\Lambda _{\mathbf{g}}\oplus
\Pi ^{1,1}$ & $\alpha _{0}=k-\theta $ & $\cdots \underset{\alpha _{i}}{\circ 
}-\underset{\alpha _{0}}{\bullet }$ & $E_{8}^{(0)}$ \\ 
$\mathbf{g}_{-1}$ & $\Lambda _{\mathbf{g}_{-1}}=\Lambda _{\mathbf{g}%
_{0}}\oplus \Pi ^{1,1}$ & $\alpha _{-1}=-\left( k+\bar{k}\right) $ & $\cdots 
\underset{\alpha _{i}}{\circ }-\underset{\alpha _{0}}{\circ }-\underset{%
\alpha _{-1}}{\bullet }$ & $E_{8}^{(1)}\equiv E_{10}$ \\ 
$\mathbf{g}_{-2}$ & $\Lambda _{\mathbf{g}_{-2}}=\Lambda _{\mathbf{g}%
_{-1}}\oplus \Pi ^{1,1}$ & $\alpha _{-2}=k-\left( \ell +\bar{\ell}\right) $
& $\cdots \underset{\alpha _{i}}{\circ }-\underset{\alpha _{0}}{\circ }-%
\underset{\alpha _{-1}}{\circ }-\underset{\alpha _{-2}}{\bullet }$ & $%
E_{8}^{(2)}\equiv E_{11}$%
\end{tabular}
\end{center}

\noindent Table 1: Extended, over extended and very extended Lie algebras,
root lattices, extensions and partial Dynkin diagrams.

To study the decomposition of the algebras we require the explicit forms and
some properties of the fundamental weights. First we report them for the
extended, over extended and very extended Lie algebras. Denoting the
fundamental weight vectors of the semisimple Lie algebra $\mathbf{g}$ as $%
\lambda _{i}^{f}$, $i=1,\ldots ,r$, the authors of \cite{gaberdiel2002class}
constructed the fundamental weights for the over extended and very extended
algebras as%
\begin{eqnarray}
\lambda _{i}^{o} &=&\lambda _{i}^{f}+n_{i}\lambda _{0}^{o},~~\ \lambda
_{0}^{o}=\bar{k}-k,\quad \lambda _{-1}^{o}=-k,  \label{weights} \\
\lambda _{i}^{v} &=&\lambda _{i}^{f}+n_{i}\lambda _{0}^{v}~~\ \lambda
_{0}^{v}=\bar{k}-k+\frac{\ell +\bar{\ell}}{2},\quad \lambda
_{-1}^{v}=-k,\quad \lambda _{-2}^{v}=-\frac{\ell +\bar{\ell}}{2},~~~~
\end{eqnarray}%
respectively, with $i=1,\ldots ,r$. Using an inner product of Lorentzian
type as defined in (\ref{ip}), these weights satisfy the orthogonality
relations%
\begin{eqnarray}
\lambda _{i}^{o}\cdot \alpha _{j} &=&\delta _{ij},\qquad i,j=-1,0,1,\ldots
,r,~~~\ \ \ \text{ \ \ }  \label{delta} \\
\lambda _{i}^{v}\cdot \alpha _{j} &=&\delta _{ij},\qquad
i,j=-2,-1,0,1,\ldots ,r,
\end{eqnarray}%
with $\alpha _{i}$ being simple roots of the appropriate root spaces. The
corresponding \emph{Weyl vectors} $\rho $, defined as the sum over all
fundamental roots are then obtained as \cite{goddard1985algebras}%
\begin{eqnarray}
\rho ^{o} &=&\sum_{j=-1}^{r}\lambda _{j}=\rho ^{f}+h\bar{k}-(1+h)k,
\label{W1} \\
\rho ^{v} &=&\sum_{j=-2}^{r}\lambda _{j}=\rho ^{f}+h\bar{k}-(1+h)k-(1-h)%
\frac{\ell +\bar{\ell}}{2},  \label{W2}
\end{eqnarray}%
respectively, with $h$ denoting the \emph{Coxeter number} and $\rho ^{f}$
the Weyl vector of the finite dimensional semisimple Lie algebra.

\section{N-extended root lattices, weight lattices and Weyl vectors}

Let us now enlarge these systems further and define the extended algebras $%
\mathbf{g}_{-n}$ with root lattices comprised of the root lattice $\Lambda _{%
\mathbf{g}_{0}}$ of the rank $r$ semisimple Lie algebra $\mathbf{g}$
extended by $n$ copies of the self-dual Lorentzian lattice $\Pi ^{1,1}$%
\begin{equation}
\Lambda _{\mathbf{g}_{-n}}=\Lambda _{\mathbf{g}}\oplus \Pi _{1}^{1,1}\oplus
\ldots \oplus \Pi _{n}^{1,1}.
\end{equation}%
Each of the root spaces $\Pi _{i}^{1,1}$, $i=1,\ldots ,n$ is equipped with
two null vectors $k_{i}$ and $\bar{k}_{i}$ with $k_{i}\cdot k_{i}=\bar{k}%
_{i}\cdot \bar{k}_{i}=0$, $k_{i}\cdot \bar{k}_{i}=1$ and two vectors $\pm
\left( k_{i}+\bar{k}_{i}\right) $ of length $2$. The simple root systems
then consists of the $r$ simple roots $\alpha _{i}~$of the semisimple Lie
algebra $\mathbf{g}$, the modified affine root $\alpha _{0}$ and $n$
extended roots $\alpha _{-i}$, $i=1,\ldots ,n$ 
\begin{equation}
\alpha ^{(n)}:=\left\{ \alpha _{1},\ldots ,\alpha _{r},\alpha
_{0}=k_{1}-\theta ,\alpha _{-1}=-\left( k_{1}+\bar{k}_{1}\right) ,\ldots
,\alpha _{-j}=k_{j-1}-\left( k_{j}+\bar{k}_{j}\right) \right\} ,
\label{roots}
\end{equation}%
for $j=2,\ldots ,n$. Using the orthogonality relation%
\begin{equation}
\lambda _{i}^{(n)}\cdot \alpha _{j}^{(n)}=\delta _{ij},\qquad
i,j=-n,0,1,\ldots ,r,  \label{ortho}
\end{equation}%
together with $\lambda _{i}^{(n)}=\sum_{j=-n}^{r}K_{ij}^{-1}\alpha
_{j}^{(n)} $, $K_{ij}^{-1}=\lambda _{i}^{(n)}\cdot \lambda _{j}^{(n)}$, we
construct the $n+r+1$ fundamental weights $\lambda _{i}^{(n)}$ of the $n$%
-extended weight lattice $\Lambda _{\mathbf{g}_{-n}}$ as%
\begin{eqnarray}
\lambda _{i}^{(n)} &=&\lambda _{i}^{f}+n_{i}\lambda _{0}^{(n)},~~\ \
i=1,\ldots ,r,  \label{l1} \\
\lambda _{0}^{(n)} &=&\bar{k}_{1}-k_{1}+\frac{1}{n}\sum\limits_{i=2}^{n}%
\left[ k_{i}+(n+1-i)\bar{k}_{i}\right] ,  \label{l0} \\
\lambda _{-1}^{(n)} &=&-k_{1}, \\
\lambda _{-2}^{(n)} &=&-\frac{1}{n}\sum\limits_{i=2}^{n}\left[ k_{i}+(n+1-i)%
\bar{k}_{i}\right] , \\
\lambda _{-3}^{(n)} &=&\frac{1}{n}(n-2)(k_{2}-\bar{k}_{2})-\frac{2}{n}%
\sum\limits_{i=3}^{n}\left[ k_{i}+(n+1-i)\bar{k}_{i}\right] ,
\end{eqnarray}%
\begin{eqnarray}
\lambda _{-4}^{(n)} &=&\frac{1}{n}(n-3)(k_{2}-\bar{k}_{2}+k_{3}-2\bar{k}%
_{3})-\frac{3}{n}\sum\limits_{i=4}^{n}\left[ k_{i}+(n+1-i)\bar{k}_{i}\right]
, \\
&&\vdots  \notag \\
\lambda _{-\ell }^{(n)} &=&\frac{1}{n}(n+1-\ell )\sum\limits_{i=2}^{\ell -1}%
\left[ k_{i}+(1-i)\bar{k}_{i}\right] +\frac{(1-\ell )}{n}\sum\limits_{i=\ell
}^{n}\left[ k_{i}+(n+1-i)\bar{k}_{i}\right] , \\
&=&\frac{(1-\ell )}{n}\sum\limits_{i=2}^{n}\left[ k_{i}+(1-i)\bar{k}_{i}%
\right] +\sum\limits_{i=2}^{\ell -1}\left[ k_{i}+(1-i)\bar{k}_{i}\right]
+(1-\ell )\sum\limits_{i=\ell }^{n}\bar{k}_{i},  \label{ln}
\end{eqnarray}%
Summing up these weights we derive the Weyl vector for the $n$-extended
system%
\begin{eqnarray}
\rho ^{(n)} &=&\sum_{j=-n}^{r}\lambda _{j}  \label{rn} \\
&=&\rho ^{f}+h\bar{k}_{1}-(1+h)k_{1}+\sum\limits_{i=2}^{n}\left[ \left( 
\frac{h}{n}+\frac{n+1-2i}{2}\right) k_{i}+\frac{(n+1-i)(2h+n(1-i))}{2n}\bar{k%
}_{i}\right] .  \notag
\end{eqnarray}%
For $n=1$ and $n=2$ the expressions reduce to the previously known formulae
in (\ref{W1}) and (\ref{W2}), respectively.

Having found a generic expression for the Weyl vector $\rho ^{(n)}$, we can
now determine the generalisation of the Freudenthal-de Vries strange formula
by computing its square. For a semisimple Lie algebra $\mathbf{g}$ with rank 
$r$ it is well known to be, see e.g. \cite{Hum,burns2000} and references
therein,%
\begin{equation}
\left( \rho ^{f}\right) ^{2}=\frac{h}{12}\text{dim}\mathbf{g=}\frac{h(h+1)r}{%
12}.
\end{equation}%
A direct calculation using the expression (\ref{rn}) yields the
generalisation for the $n$-extended algebras%
\begin{equation}
\rho ^{(n)}\cdot \rho ^{(n)}=\frac{h(h+1)r+n(n^{2}-1)}{12}-\frac{h(h+n)(1+n)%
}{n},  \label{r2}
\end{equation}%
for $n\geq 1$. For $n=1$ this corresponds to the expression found in \cite%
{gaberdiel2002class} for the over extended case.

\section{SO(1,2) and SO(3) principal subalgebras}

As the class of Lorentzian Kac-Moody algebras is very large, several
attempts have been made in seeking for further properties that distinguish
between different subclasses. One such property that has turned out to be
very powerful when analysing integrable systems based on finite or affine
Kac-Moody algebras \cite{Mass2,kneipp1993}, as well as the structure of
Kac-Moody algebras themselves, is the feature of possessing a principal $%
SO(3)$-subalgebra \cite{Kost}. In terms of the generators in the Chevalley
basis $H_{j}$, $E_{i}$, $F_{i}$, obeying the standard commutation relations $%
[H_{i},H_{j}]$ $=0$, $[E_{i},F_{j}]$ $=\delta _{ij}H_{i}$, $[H_{i},E_{j}]$ $%
=K_{ij}E_{j}$, $[H_{i},F_{j}]$ $=-K_{ij}F_{j}$ with $K$ denoting the Cartan
matrix, the principal $SO(3)$-generators

\begin{equation}
J_{3}=\sum_{i=1}^{r}D_{i}H_{i},\quad J_{+}=\sum_{i=1}^{r}n_{i}E_{i},\quad
J_{-}=\sum_{i=1}^{r}n_{i}F_{i},\quad D_{i}:=\sum_{j=1}^{r}K_{ji}^{-1},
\label{so3}
\end{equation}%
satisfy $\left[ J_{+},J_{-}\right] =J_{3}$, $\left[ J_{3},J_{\pm }\right]
=\pm J_{\pm }$. The Hermiticity properties $E_{i}^{\dagger }=F_{i}$, $%
H_{i}^{\dagger }=H_{i}$ are inherited by the generators $J_{+}$ and $J_{-}$
as $J_{+}^{\dagger }=J_{-}$ when $n_{i}\in \mathbb{R}$. The $SO(3)$%
-commutation relation $\left[ J_{+},J_{-}\right] =J_{3}$ is satisfied when $%
n_{i}=\sqrt{D_{i}}$.

In the case of the Lorentzian Kac-Moody algebras the analogue of the $SO(3)$%
-principal subalgebra is a $SO(1,2)$-principal subalgebra \cite%
{nicolai2001dio,gaberdiel2002class} with generators 
\begin{equation}
\hat{J}_{3}=-\sum_{i=1}^{r}\hat{D}_{i}H_{i},\quad \hat{J}_{+}=%
\sum_{i=1}^{r}p_{i}E_{i},\quad \hat{J}_{-}=\sum_{i=1}^{r}q_{i}F_{i},\quad 
\hat{D}_{i}:=\sum_{j=1}^{r}K_{ji}^{-1},
\end{equation}%
satisfying $\left[ \hat{J}_{+},\hat{J}_{-}\right] =-\hat{J}_{3}$, $\left[ 
\hat{J}_{3},\hat{J}_{\pm }\right] =\pm \hat{J}_{\pm }$ and being Hermitian
when\ $p_{i}q_{i}=\left\vert p_{i}\right\vert ^{2}=-\hat{D}_{i}$. Thus \emph{%
a necessary and sufficient condition for the existence of a }$SO(3)$\emph{%
-principal subalgebra or a }$SO(1,2)$\emph{-principal subalgebra is }$%
D_{i}>0 $\emph{\ or }$\hat{D}_{i}<0$\emph{\ for all }$i$\emph{, respectively}%
.

We argue further that a necessary condition for an extended algebra $\mathbf{%
g}_{-n}$ to possess a $SO(3)$-principal subalgebra and a $SO(1,2)$-principal
subalgebra is that there exists a $k\in S=\{-n,\ldots ,0,1,\ldots ,r\}$ such
that $D_{k}=\sum_{j=-n}^{r}K_{jk}^{-1}=0$. We may then decompose the index
set $S$ as $\tilde{S}=S\backslash \{k\}=S_{1}\cup S_{2}$, such that $%
K_{ij}=0 $ for all $i\in S_{1}$, $j\in S_{2}$ and $K_{i^{\prime }k}\neq 0$, $%
K_{j^{\prime }k}\neq 0$ for two specific $i^{\prime }\in S_{1}$ and $%
j^{\prime }\in S_{2}$. Thus removing the node $k$ from the connected Dynkin
diagram $\mathbf{g}_{-n}$ will decompose it into two connected diagrams such
that two generators indexed by $i\in S_{1}$ and $j\in S_{2}$ will commute.
Thus when $D_{i}>0$ for $i\in S_{1}$ and $D_{j}<0$ for $j\in S_{2}$ we can
formulate two commuting principal subalgebras with generators $%
\{J_{3},J_{\pm }\}$ and $\{\hat{J}_{3},\hat{J}_{\pm }\}$. For instance, we
have%
\begin{equation}
\left[ J_{3},\hat{J}_{+}\right] =\sum\nolimits_{i\in S_{1},j\in S_{2}}D_{i}%
\sqrt{-\hat{D}_{j}}\left[ H_{i},E_{j}\right] =\sum\nolimits_{i\in S_{1},j\in
S_{2}}D_{i}\sqrt{-\hat{D}_{j}}K_{ij}E_{j}=0,
\end{equation}%
and similarly for the other generators. This commuting structure extends to
the $SO(3)$ and $SO(1,2)$ Casimir operators%
\begin{equation}
C=J_{3}J_{3}-J_{+}J_{-}-J_{-}J_{+}\text{,\quad and\quad }\hat{C}=\hat{J}_{3}%
\hat{J}_{3}-\hat{J}_{+}\hat{J}_{-}-\hat{J}_{-}\hat{J}_{+}\text{,}
\end{equation}%
respectively. So that we have $SO(3)\oplus SO(1,2)$ with $[C,\hat{C}]=0\,$.

Computing the inner products of the generators in the adjoint
representation, as carried out for instance in \cite{nicolai2001dio}, yields%
\begin{eqnarray}
(J_{3},J_{3}) &=&\rho ^{(n)}\cdot \rho ^{(n)}>0,\qquad \text{and\qquad }%
(J_{\pm },J_{\pm })=\rho ^{(n)}\cdot \rho ^{(n)}>0,  \label{s3} \\
(\hat{J}_{3},\hat{J}_{3}) &=&\rho ^{(n)}\cdot \rho ^{(n)}<0,\qquad \text{%
and\qquad }(\hat{J}_{\pm },\hat{J}_{\pm })=-\rho ^{(n)}\cdot \rho ^{(n)}>0.
\label{s4}
\end{eqnarray}%
Thus identifying the signatures of $\rho ^{(n)}\cdot \rho ^{(n)}$ serves as
a necessary condition for the existence of the respective principal
subalgebras. Using the generalised Freudenthal-de Vries strange formula (\ref%
{r2}) for a given semisimple Lie algebra $\mathbf{g}$ with rank $r$, the
maximal value of $n_{\text{max}}$ for $\mathbf{g}_{-n}$ to possess a $%
SO(1,2) $-principal subalgebras is easily determined from the inequalities
in (\ref{s4}). Using relation (\ref{r2}) with a given rank we compute for
the exceptional semisimple Lie algebras%
\begin{equation}
E_{6}:\text{\thinspace }n_{\text{max}}=23,\qquad E_{7}:\text{\thinspace }n_{%
\text{max}}=17,\qquad E_{8}:\text{\thinspace }n_{\text{max}}=14.
\end{equation}%
For the $A_{r}$ and $D_{r}$ algebras we present the results in figure \ref%
{Fig1} for different values of $r$.

\FIGURE{ \epsfig{file=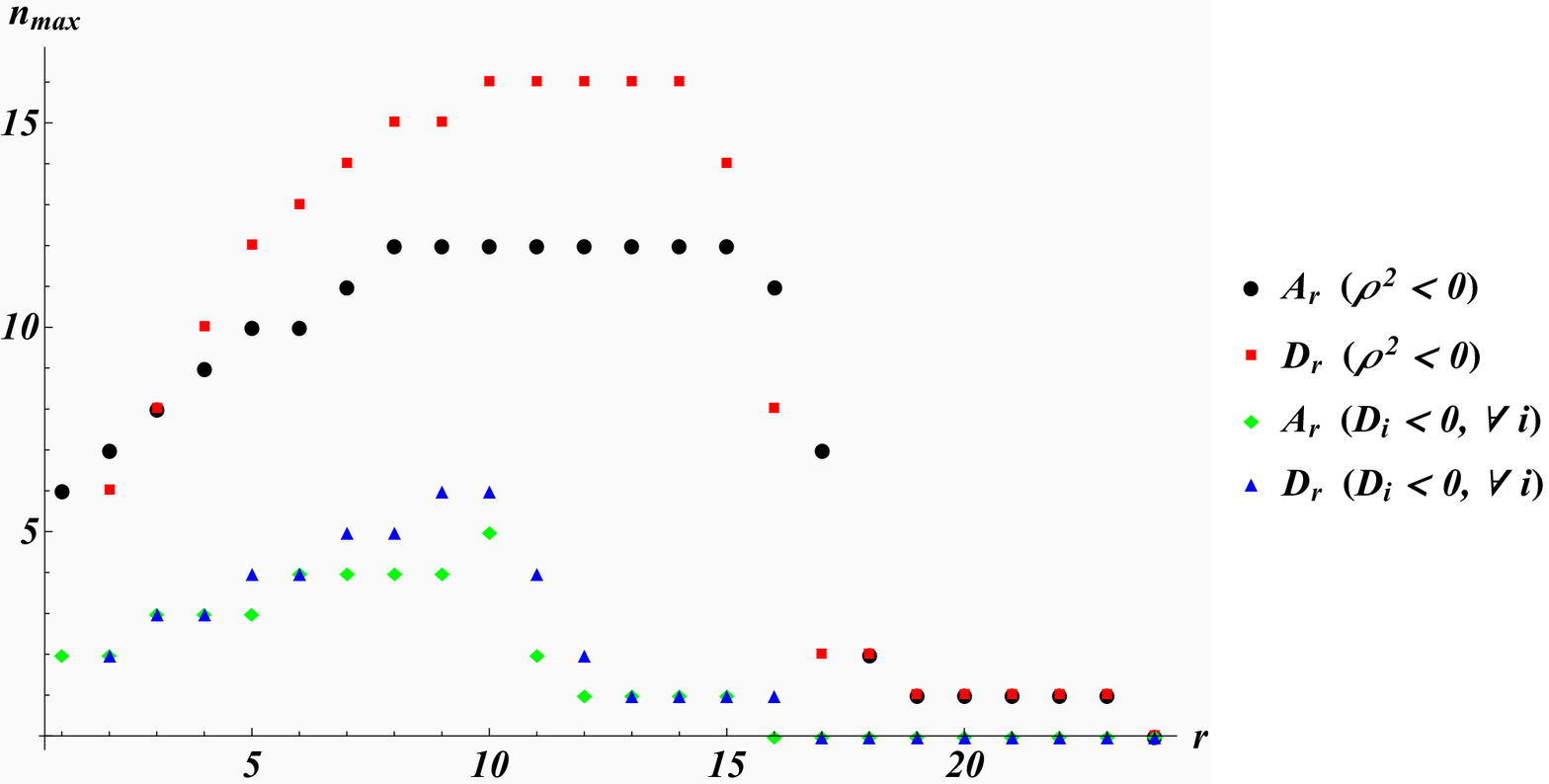, height=8cm} 
\caption{Maximum values of $n$ for $\mathbf{g}_{-n}$ with rank $r$ to possess a SO(1,2)-principal subalgebra from the necessary condition $\rho ^{2}<0$ versus
the necessary and sufficient condition $D_{i}<0$, $\forall i$.}
       \label{Fig1}}

We observe from figure \ref{Fig1} that for $A_{r}$ and $D_{r}$ with $r\geq
24 $ no $n$--extended algebra exists that possesses a $SO(1,2)$-principal
subalgebras. This agrees with the findings in \cite{goddard1986kac} for $n=1$%
. For $r<24$ such a possibility exists, but $\rho ^{2}<0$ implies it does
not exist when $n>12$ and $n>16$, for $A_{r}$ and $D_{r}$, respectively%
\footnote{%
For the over and very extended cases our results differ mildly in one case
from a typo in \cite{gaberdiel2002class}, where it was stated that also the
over extended $A_{16}^{(1)}$ possess a $SO(1,2)$-principal subalgebras.}. As
the criterion (\ref{s4}) is only necessary but not sufficient, let us
compute the values for $D_{i}^{(n)}$ to obtain the more restrictive
necessary and sufficient information.

\section{Expansion coefficients of the diagonal principal subalgebra
generator}

Having constructed the expressions for all fundamental weights and the Weyl
vector, we can evaluate the expansion coefficients $D_{i}$ directly from the
definition (\ref{so3}). Focussing here on the case for which the finite
semisimple part is simply laced, so that all roots have length $2$, the
inverse Cartan matrix is symmetric and acquires a simple form in terms of
the fundamental weights $\lambda _{i}^{(n)}$ as $K_{ji}^{-1}=$ $\lambda
_{j}^{(n)}\cdot \lambda _{i}^{(n)}$. Therefore the constants 
\begin{equation}
D_{k}^{(n)}=\sum_{j=-n}^{r}K_{kj}^{-1}=\rho ^{(n)}\cdot \lambda
_{k}^{(n)},~~~\ \ \ k=-n,\ldots ,-1,0,1,\ldots ,r,  \label{Di}
\end{equation}%
can be computed either by using the generic expressions for the weight
vectors (\ref{l1})-(\ref{ln}) and Weyl vectors (\ref{rn}) or by directly
inverting the Cartan matrix as in (\ref{Di}). From the generic expressions
we derive general formulae for the expansion coefficients%
\begin{equation}
D_{i}^{(n)}=D_{i}^{f}+n_{i}D_{0}^{(n)},\quad \text{and\quad }%
D_{-j}^{(n)}=(n-j+1)\left( \frac{j-1}{2}-\frac{h}{n}\right) ,~~i=1,\ldots
,r;j=0,\ldots ,n,  \label{genD}
\end{equation}%
for the semisimple Lie algebraic and extended part, respectively. We
abbreviated $D_{i}^{f}:=\rho ^{f}\cdot \lambda _{i}^{f}$. For the over
extended and very extended algebras the expressions in (\ref{genD}) become
for instance%
\begin{eqnarray}
D_{-1}^{o} &=&-h,\quad ~~D_{0}^{o}=-(2h+1),\quad
~~D_{i}^{o}=D_{i}^{f}+n_{i}D_{0}^{o},~~ \\
D_{-2}^{v} &=&\frac{1}{2}(1-h),\quad ~~D_{-1}^{v}=-h,\quad ~~D_{0}^{v}=-%
\frac{3}{2}(h+1),\quad ~~D_{i}^{v}=D_{i}^{f}+n_{i}D_{0}^{v}.~~~
\end{eqnarray}%
The fundamental Weyl vectors $\rho ^{f}$, Coxeter numbers $h$ and Kac labels 
$n_{i}$ are algebra specific and well known, see e.g. \cite%
{bourbaki1968groupes}. We list them here for convenience in table 2.

\begin{center}
\begin{tabular}{l||l|l|l}
& Kac labels $n_{i}$ & exponents $e_{i}$ & Coxeter number $h$ \\ \hline\hline
$A_{r}$ & $1,\ldots ,1$ & $1,2,3,\ldots ,r$ & $r+1$ \\ \hline
$D_{r}$ & $1,2,2,\ldots ,2,1,1$ & $1,3,5,\ldots ,2r-5,2r-3,r-1$ & $2r-2$ \\ 
\hline
$E_{6}$ & $1,2,2,3,2,1$ & $1,4,5,7,8,11$ & $12$ \\ \hline
$E_{7}$ & $2,2,3,4,3,2,1$ & $1,5,7,9,11,13,17$ & $18$ \\ \hline
$E_{8}$ & $2,3,4,6,5,4,3,2$ & $1,7,11,13,17,19,23,29$ & $30$%
\end{tabular}

Table 2: Kac labels, exponents and Coxeter number for the simply laced Lie
algebras.
\end{center}

\noindent Also the Weyl vectors are known in these cases in terms of the
simple roots%
\begin{eqnarray}
A_{r} &:&~\rho ^{f}=\sum\limits_{i=1}^{r}\frac{i}{2}(r-i+1)\alpha _{i}\,,%
\text{~~} \\
D_{r} &:&~\rho ^{f}=\sum\limits_{i=1}^{r-2}\left[ ir-\frac{i(i+1)}{2}\right]
\alpha _{i}+\frac{r(r-1)}{4}(\alpha _{r-1}+\alpha _{r})\text{,~} \\
E_{6} &:&\rho ^{f}=(8\alpha _{1}+11\alpha _{2}+15\alpha _{3}+21\alpha
_{4}+15\alpha _{5}+8\alpha _{6}\text{,~~} \\
E_{7} &:&\rho ^{f}=\frac{1}{2}(34\alpha _{1}+49\alpha _{2}+66\alpha
_{3}+96\alpha _{4}+75\alpha _{5}+52\alpha _{6}+27\alpha _{7})\text{,~~} \\
E_{8} &:&\rho ^{f}=46\alpha _{1}+68\alpha _{2}+91\alpha _{3}+135\alpha
_{4}+110\alpha _{5}+84\alpha _{6}+57\alpha _{7}+29\alpha _{8}\text{.}~~~\ 
\end{eqnarray}

\noindent Hence we compute the terms $D_{i}^{f}$ in the general expressions
for the expansion coefficients (\ref{genD}) as 
\begin{eqnarray}
A_{r} &:&D_{i}^{f}=\frac{i}{2}(r-i+1),~~i=1,\ldots ,r \\
D_{r} &:&D_{r-1}^{f}=D_{r}^{f}=\frac{r(r-1)}{4}\,,~D_{j}^{f}=jr-\frac{j(j+1)%
}{2},~~j=1,\ldots ,r-2  \notag \\
E_{6}
&:&D_{1}^{f}=8,~D_{2}^{f}=11,~D_{3}^{f}=15,~D_{4}^{f}=21,~D_{5}^{f}=15,~D_{6}^{f}=8,
\notag \\
E_{7} &:&D_{1}^{f}=17,~D_{2}^{f}=\frac{49}{2}%
,~D_{3}^{f}=33,~D_{4}^{f}=48,~D_{5}^{f}=\frac{75}{2}%
,~D_{6}^{f}=26,~D_{7}^{f}=\frac{27}{2},  \notag \\
E_{8} &:&\
D_{1}^{f}=46,~D_{2}^{f}=68,~D_{3}^{f}=91,~D_{4}^{f}=135,~D_{5}^{f}=110,~D_{6}^{f}=84,~D_{7}^{f}=57,~D_{8}^{f}=29.
\notag
\end{eqnarray}%
Evidently all constants $D_{i}^{f}$ for all semisimple Lie algebras are
positive. For the over extended algebras we obtain therefore 
\begin{eqnarray}
A_{r}^{(1)} &:&D_{-1}^{o}=-(r+1),~~~~D_{0}^{o}=-(2r+3),~~\ D_{i}^{o}=\frac{i%
}{2}(r-i+1)-(2r+3)~~  \label{D1} \\
D_{r}^{(1)}
&:&D_{-1}^{o}=2-2r,~D_{0}^{o}=3-4r,~D_{1}^{o}=2-3r,~D_{j}^{o}=(j-8)r-\frac{%
j(j+1)}{2}+6, \\
&&D_{r-1}^{o}=D_{r}^{o}=\frac{r(r+1)}{4}+3-4r,  \notag \\
E_{6}^{(1)}
&:&D_{-1}^{o}=-12,~D_{0}^{o}=-25,~D_{1}^{o}=-17,~D_{2}^{o}=-89,~D_{3}^{o}=-110,~D_{4}^{o}=-154,
\\
&&D_{5}^{o}=-185,~D_{6}^{o}=-267,  \notag \\
E_{7}^{(1)} &:&D_{-1}^{o}=-18,~D_{0}^{o}=-37,~D_{1}^{o}=-57,~D_{2}^{o}=-%
\frac{99}{2},~D_{3}^{o}=-78,~D_{4}^{o}=-100,~ \\
&&D_{5}^{o}=-\frac{147}{2},~D_{6}^{o}=-48,~D_{7}^{o}=-\frac{47}{2},  \notag
\\
E_{8}^{(1)}
&:&D_{-1}^{o}=-30,~D_{0}^{o}=-61,~D_{1}^{o}=-15,~D_{2}^{o}=-359,~D_{3}^{o}=-580,~D_{4}^{o}=-658,~~~~
\\
&&D_{5}^{o}=-927,~D_{6}^{o}=-1075,~D_{7}^{o}=-1346,~D_{8}^{o}=-1740,  \notag
\end{eqnarray}%
with $i=1,\ldots ,n$, $j=2,\ldots ,n-2$ and for the very extended algebras
we compute%
\begin{eqnarray}
A_{r}^{(2)} &:&D_{-2}^{v}=-\frac{r}{2},~D_{-1}^{v}=-(r+1),~D_{0}^{v}=-\frac{3%
}{2}(r+2),~D_{i}^{v}=\frac{r}{2}(i-3)+\frac{i}{2}(1-i)-3,~~~~\ \ \ 
\label{D2} \\
D_{r}^{(2)} &:&D_{-2}^{v}=\frac{3}{2}-r,~D_{-1}^{v}=2-2r,~D_{0}^{v}=\frac{3}{%
2}-3r,~D_{1}^{v}=\frac{1}{2}-2r, \\
&&D_{j}^{v}=(j-6)r-\frac{j(j+1)}{2}+3,~D_{r-1}=D_{r}=\frac{r(r+1)}{4}+\frac{3%
}{2}-3r,  \notag
\end{eqnarray}%
\begin{eqnarray}
E_{6}^{(2)} &:&D_{-2}^{v}=-\frac{11}{2},~D_{-1}^{v}=-12,~D_{0}^{v}=-\frac{39%
}{2},~D_{1}^{v}=D_{6}^{v}=-\frac{23}{2},~D_{2}^{v}=-28,~ \\
&&D_{3}^{v}=D_{5}^{v}=-24,~D_{4}^{v}=-\frac{75}{2},  \notag \\
E_{7}^{(2)} &:&D_{-2}^{v}=-\frac{17}{2},~D_{-1}^{v}=-18,~D_{0}^{v}=-\frac{57%
}{2},~D_{1}^{v}=-40,~D_{2}^{v}=-\frac{65}{2},~D_{3}^{v}=-\frac{105}{2}%
,~~~~~~~~~ \\
&&D_{4}^{v}=-66,~D_{5}^{v}=-48,~D_{6}^{v}=-31,~D_{7}^{v}=-15,  \notag \\
E_{8}^{(2)} &:&D_{-2}^{v}=-\frac{29}{2},~D_{-1}^{v}=-30,~D_{0}^{v}=-\frac{93%
}{2},~D_{1}^{v}=-47,~D_{2}^{v}=-\frac{143}{2},~D_{3}^{v}=-95, \\
&&D_{4}^{v}=-144,~D_{5}^{v}=-\frac{245}{2}x,~D_{6}^{v}=-102,~D_{7}^{v}=-%
\frac{165}{2},D_{8}^{v}=-64.  \notag
\end{eqnarray}%
with $i=1,\ldots ,n$, $j=2,\ldots ,n-2$.

From these expression we find directly the maximal value of $n$ for $\mathbf{%
g}_{-n}$ with rank $r$ to possess a $SO(1,2)$-principal subalgebra from the
necessary and sufficient condition $D_{i}<0$, $\forall i$. For the
exceptional Lie algebras we obtain%
\begin{equation}
E_{6}:\text{\thinspace }n_{\text{max}}=5,\qquad E_{7}:\text{\thinspace }n_{%
\text{max}}=6,\qquad E_{8}:\text{\thinspace }n_{\text{max}}=7.
\end{equation}%
For $A_{r}$ and $D_{r}$ the results are reported in figure \ref{Fig1}.
Comparing these exact values to those resulting from the analysis of the
necessary condition $\rho ^{2}<0$ shows consistency, but also that the
latter values are much more restrictive.

\section{Direct decomposition of $n$-extended Lorentzian Kac-Moody algebras}

As argued above, when a constant $D_{k}^{(n)}$ vanishes we can potentially
simultaneously find a $SO(1,2)$-principal subalgebras and a $SO(3)$%
-principal subalgebra. This requires, however, that the $D_{i}^{(n)}$ for $i$
belonging to the two separate index sets $S_{1}$ and $S_{2}$ are of definite
sign. If that is not the case the algebra can be decomposed further. To
identify when either of these scenarios occurs, we can set our solutions in (%
\ref{genD}) to zero and solve for $n,i,j$, with the only meaningful
solutions being those for which $n,i\in \mathbb{N}$ and $i\leq n$, $j\leq n$.

For the extended parts of the Dynkin diagrams we easily find from (\ref{genD}%
)%
\begin{equation}
D_{-j}^{(n)}=0,~~~~~~~\ \text{for }j=1+\frac{2h}{n}.
\end{equation}%
For a given value of $n$ there can only be a finite number of solutions due
to the restriction $j\leq n$. Using the Coxeter numbers from table 2, we
find the solutions 
\begin{eqnarray}
A_{r}^{(n)} &:&D_{i}^{(n)}=0~~~~\ \text{for }%
(n,r,j)=(3,2,3),(4,1,2),(4,3,2),(4,5,4),(5,4,3),\ldots \\
D_{r}^{(n)} &:&D_{i}^{(n)}=0~~~~\ \text{for }%
(n,r,j)=(4,4,4),(5,6,5),(6,4,3),(6,7,5),(7,8,5),\ldots  \label{dx}
\end{eqnarray}%
and for the exceptional Lie algebras the only possible solutions are%
\begin{eqnarray}
E_{6}^{(n)} &=&E_{6}^{(j-1)}\diamond L\diamond A_{n-j}~~\text{for }%
(n,j)=(6,5),(8,4),(12,3),(24,2), \\
E_{7}^{(n)} &=&E_{7}^{(j-1)}\diamond L\diamond A_{n-j}~~\text{for }%
(n,j)=(9,5),(12,4),(18,3),(36,2), \\
E_{8}^{(n)} &=&E_{8}^{(j-1)}\diamond L\diamond A_{n-j}~~\text{for }%
(n,j)=(10,7),(12,6),(15,5),(20,4),(30,3),(60,2).~~~~
\end{eqnarray}%
We denote the Lorentzian root corresponding to the node that needs to be
deleted by $L$.

For the parts of the Dynkin diagrams corresponding to semisimple Lie
algebras also the expressions for $\rho ^{f}$ need to be treated
case-by-case. We find%
\begin{eqnarray}
A_{r}^{(n)} &:&D_{i}^{(n)}=0~~~~\ \text{for ~}i=\frac{r+1}{2}\pm \frac{1}{2}%
\sqrt{r^{2}-6r-4n-11-\frac{8(1+r)}{n}}, \\
D_{r}^{(n)} &:&D_{i}^{(n)}=0~~~~\ \text{for ~}i=\frac{r-1}{2}\pm \sqrt{%
r^{2}-9r-2n+\frac{25}{4}+\frac{8(1-r)}{n}}.
\end{eqnarray}%
For the over and very extended algebras the only solutions are%
\begin{eqnarray}
A_{r}^{(1)} &:&\text{~}r=16,i=7,10\text{; ~}r=18,i=6,13\text{; ~}r=26,j=5,22%
\text{,} \\
A_{r}^{(2)} &:&\text{~}r=12,i=6,7\text{; ~}r=13,i=5,9\text{; ~}r=18,j=4,15%
\text{,} \\
D_{r}^{(1)} &:&\text{~}r=17,j=13\text{; ~}r=18,j=12\text{; ~}r=20,j=11\text{%
;~~}r=39,j=9\text{,}  \notag \\
D_{r}^{(2)} &:&\text{~}r=13,j=10\text{; ~}r=14,j=9\text{; ~}r=25,j=7\text{.}
\notag
\end{eqnarray}%
There are no solutions for the $E$-series on this part of the Dynkin
diagram. The complete list of solution with corresponding decomposition is
presented in the tables 3 and 4. We observe that in the reduced part we also
obtain some algebras that are not of the $n$-extended form as described
above. To refer to them we introduce the notation $\hat{A}_{r}^{(n,m)}$
labelling an $A_{r}$-Dynkin diagram with $n$ roots successively attached to
the $m$-th node in form of an $A_{n}$-algebra. The special case of the $n$%
-extended symmetric Dynkin diagram with $n$ roots attached to the middle
node of $A_{r}$ we denote by $\hat{A}_{r}^{(n)}$. Some of the $\hat{A}%
_{r}^{(n,m)}$-algebras are equivalent to the $n$-extended versions of the $E$%
-series. We have $\hat{A}_{5}^{(n+2,3)}\equiv E_{6}^{(n-2)}$, $\hat{A}%
_{n}^{(1,4)}\equiv E_{7}^{(n-7)}$ and $\hat{A}_{n}^{(1,3)}\equiv
E_{8}^{(n-8)}$. We also have the symmetries $\hat{A}_{r}^{(n,m)}=\hat{A}%
_{r}^{(n,r+1-m)}=\hat{A}_{n+m}^{(r+1-m,m)}=\hat{A}_{r+n+1-m}^{(m-1,m)}$. In
the resulting decomposition we also encounter algebras that decompose
further by possessing Lorentzian roots on their extended legs of the
corresponding Dynkin diagrams. We mark them in bold in tables 3 and 4. The
precise way in which they decompose is reported below in tables 6 and 7.

\begin{center}
\begin{tabular}{|l|l|l|}
\hline
$A_{16}^{(1)}=\hat{A}_{13}^{(1)}\diamond L^{2}\diamond A_{2}$ & $%
~A_{18}^{(1)}=\hat{A}_{11}^{(1)}\diamond L^{2}\diamond A_{6}$ & $%
~A_{19}^{(1)}=\hat{A}_{11}^{(1)}\diamond L^{2}\diamond A_{16}$ \\ \hline
$A_{12}^{(2)}=\hat{A}_{11}^{(2)}\diamond L^{2}$ & $A_{13}^{(2)}=\hat{A}%
_{9}^{(2)}\diamond L^{2}\diamond A_{3}$ & $A_{18}^{(2)}=\hat{A}%
_{7}^{(2)}\diamond L^{2}\diamond A_{10}$ \\ \hline
$A_{12}^{(3)}=\hat{A}_{11}^{(3)}\diamond L$ & $A_{14}^{(3)}=\hat{A}%
_{7}^{(3)}\diamond L^{2}\diamond A_{6}$ & $A_{38}^{(3)}=E_{6}^{(1)}\diamond
L^{2}\diamond A_{32}$ \\ \hline
$A_{11}^{(4)}=\hat{A}_{9}^{(4)}\diamond L^{2}\diamond A_{1}$ & $A_{13}^{(4)}=%
\hat{A}_{7}^{(4)}\diamond L^{2}\diamond A_{5}$ & $A_{27}^{(4)}=E_{6}^{(2)}%
\diamond L^{2}\diamond A_{21}$ \\ \hline
$A_{24}^{(5)}=E_{6}^{(3)}\diamond L^{2}\diamond A_{18}$ & $A_{11}^{(6)}=%
\mathbf{\hat{A}}_{\mathbf{9}}^{\mathbf{(6)}}\diamond L^{2}\diamond A_{1}$ & $%
A_{23}^{(6)}=E_{6}^{(4)}\diamond L^{2}\diamond A_{17}$ \\ \hline
$A_{13}^{(7)}=\mathbf{\hat{A}}_{\mathbf{7}}^{\mathbf{(7)}}\diamond
L^{2}\diamond A_{5}$ & $A_{11}^{(8)}=\mathbf{\hat{A}}_{\mathbf{11}}^{\mathbf{%
(8)}}\diamond L$ & $A_{23}^{(8)}=\mathbf{E}_{\mathbf{6}}^{\mathbf{(6)}%
}\diamond L^{2}\diamond A_{17}$ \\ \hline
$A_{14}^{(10)}=\mathbf{\hat{A}}_{\mathbf{7}}^{\mathbf{(10)}}\diamond
L^{2}\diamond A_{6}$ & $A_{24}^{(10)}=\mathbf{E}_{\mathbf{6}}^{\mathbf{(8)}%
}\diamond L^{2}\diamond A_{18}~~$ & $A_{12}^{(13)}=\mathbf{\hat{A}}_{\mathbf{%
11}}^{\mathbf{(13)}}\diamond L^{2}$ \\ \hline
$A_{13}^{(14)}=\mathbf{\hat{A}}_{\mathbf{9}}^{\mathbf{(14)}}\diamond
L^{2}\diamond A_{3}$ & $A_{27}^{(14)}=\mathbf{E}_{6}^{\mathbf{(12)}}\diamond
L^{2}\diamond A_{21}$ & $A_{18}^{(19)}=\mathbf{\hat{A}}_{\mathbf{7}}^{%
\mathbf{(19)}}\diamond L^{2}\diamond A_{10}$ \\ \hline
$A_{38}^{(26)}=\mathbf{E}_{6}^{\mathbf{(24)}}\diamond L^{2}\diamond A_{32}~~$
& $A_{16}^{(34)}=\mathbf{\hat{A}}_{\mathbf{13}}^{\mathbf{(34)}}\diamond
L^{2}\diamond A_{2}$ & $A_{18}^{(38)}=\mathbf{\hat{A}}_{\mathbf{11}}^{%
\mathbf{(38)}}\diamond L^{2}\diamond A_{6}$ \\ \hline
$A_{26}^{(54)}=\mathbf{\hat{A}}_{\mathbf{9}}^{\mathbf{(54)}}\diamond
L^{2}\diamond A_{16}$ &  &  \\ \hline
\end{tabular}

Table 3: Decomposition of the $n$-extended algebras $A_{r}^{(n)}$.

\begin{tabular}{|l|l|l|}
\hline
$D_{17}^{(1)}=E_{8}^{(5)}\diamond L\diamond D_{4}$ & $%
D_{18}^{(1)}=E_{8}^{(4)}\diamond L\diamond D_{6}$ & $%
~D_{20}^{(1)}=E_{8}^{(3)}\diamond L\diamond D_{9}$ \\ \hline
$D_{39}^{(1)}=E_{8}^{(1)}\diamond L\diamond D_{30}$ & $%
D_{13}^{(2)}=E_{7}^{(4)}\diamond L\diamond A_{3}$ & $%
D_{14}^{(2)}=E_{7}^{(3)}\diamond L\diamond D_{5}$ \\ \hline
$D_{25}^{(2)}=E_{7}^{(1)}\diamond L\diamond D_{18}$ & $D_{13}^{(3)}=\hat{A}%
_{10}^{(1,5)}\diamond L\diamond D_{5}$ & $D_{16}^{(3)}=\hat{A}%
_{9}^{(1,5)}\diamond L\diamond D_{9}$ \\ \hline
$D_{11}^{(4)}=\hat{A}_{13}^{(1,6)}\diamond L^{2}$ & $D_{12}^{(4)}=\hat{A}%
_{11}^{(1,6)}\diamond L\diamond D_{4}$ & $D_{14}^{(4)}=\hat{A}%
_{10}^{(1,6)}\diamond L\diamond D_{7}$ \\ \hline
$D_{21}^{(4)}=E_{7}^{(2)}\diamond L\diamond D_{15}$ & $D_{11}^{(5)}=\hat{A}%
_{13}^{(1,7)}\diamond L\diamond A_{1}^{2}$ & $D_{81}^{(5)}=E_{8}^{(1)}%
\diamond L\diamond D_{76}$ \\ \hline
$D_{13}^{(6)}=\hat{A}_{12}^{(1,5)}\diamond L\diamond D_{6}$ & $%
D_{52}^{(6)}=E_{8}^{(2)}\diamond L\diamond D_{47}$ & $%
D_{43}^{(7)}=E_{8}^{(3)}\diamond L\diamond D_{38}$ \\ \hline
$D_{11}^{(8)}=\mathbf{\hat{A}}_{\mathbf{16}}^{\mathbf{(1,7)}}\diamond
L\diamond A_{1}^{2}$ & $D_{13}^{(8)}=\mathbf{\hat{A}}_{\mathbf{14}}^{\mathbf{%
(1,5)}}\diamond L\diamond D_{6}$ & $D_{17}^{(8)}=E_{7}^{(6)}\diamond
L\diamond D_{11}$ \\ \hline
$D_{39}^{(8)}=E_{8}^{(4)}\diamond L\diamond D_{34}$ & $%
D_{37}^{(9)}=E_{8}^{(5)}\diamond L\diamond D_{32}$ & $D_{11}^{(10)}=\mathbf{%
\hat{A}}_{\mathbf{19}}^{\mathbf{(1,8)}}\diamond L^{2}$ \\ \hline
$D_{36}^{(10)}=E_{8}^{(6)}\diamond L\diamond D_{31}$ & $D_{12}^{(11)}=%
\mathbf{\hat{A}}_{\mathbf{18}}^{\mathbf{(1,6)}}\diamond L\diamond D_{4}$ & $%
D_{14}^{(13)}=\mathbf{\hat{A}}_{\mathbf{19}}^{\mathbf{(1,5)}}\diamond
L\diamond D_{7}$ \\ \hline
$D_{36}^{(14)}=\mathbf{E}_{\mathbf{8}}^{\mathbf{(10)}}\diamond L\diamond
D_{31}$ & $D_{13}^{(16)}=\mathbf{\hat{A}}_{\mathbf{23}}^{\mathbf{(1,6)}%
}\diamond L\diamond D_{5}$ & $D_{37}^{(16)}=\mathbf{E}_{\mathbf{8}}^{\mathbf{%
(12)}}\diamond L\diamond D_{32}$ \\ \hline
$D_{39}^{(19)}=\mathbf{E}_{\mathbf{8}}^{\mathbf{(15)}}\diamond L\diamond
D_{34}$ & $D_{16}^{(20)}=\mathbf{\hat{A}}_{\mathbf{26}}^{\mathbf{(1,5)}%
}\diamond L\diamond D_{9}$ & $D_{21}^{(20)}=\mathbf{E}_{\mathbf{7}}^{\mathbf{%
(18)}}\diamond L\diamond D_{15}$ \\ \hline
$D_{13}^{(24)}=\mathbf{\hat{A}}_{\mathbf{33}}^{\mathbf{(1,8)}}\diamond
L\diamond A_{3}$ & $D_{43}^{(24)}=\mathbf{E}_{\mathbf{8}}^{\mathbf{(20)}%
}\diamond L\diamond D_{38}$ & $D_{14}^{(26)}=\mathbf{\hat{A}}_{\mathbf{34}}^{%
\mathbf{(1,7)}}\diamond L\diamond D_{5}$ \\ \hline
$D_{52}^{(34)}=\mathbf{E}_{\mathbf{8}}^{\mathbf{(30)}}\diamond L\diamond
D_{47}$ & $D_{25}^{(48)}=\mathbf{\hat{A}}_{\mathbf{54}}^{\mathbf{(1,5)}%
}\diamond L\diamond D_{18}$ & $D_{17}^{(64)}=\mathbf{\hat{A}}_{\mathbf{76}}^{%
\mathbf{(1,11)}}\diamond L\diamond D_{4}$ \\ \hline
$D_{81}^{(64)}=\mathbf{E}_{\mathbf{8}}^{\mathbf{(60)}}\diamond L\diamond
D_{76}$ & $D_{18}^{(68)}=\mathbf{\hat{A}}_{\mathbf{79}}^{\mathbf{(1,10)}%
}\diamond L\diamond D_{6}$ & $D_{20}^{(76)}=\mathbf{\hat{A}}_{\mathbf{86}}^{%
\mathbf{(1,9)}}\diamond L\diamond D_{9}$ \\ \hline
$D_{39}^{(152)}=\mathbf{\hat{A}}_{\mathbf{160}}^{\mathbf{(1,7)}}\diamond
L\diamond D_{30}$ &  &  \\ \hline
\end{tabular}

Table 4: Decomposition of the $n$-extended algebras $D_{r}^{(n)}$.
\end{center}

\subsection{Reduced system versus $n$-extended versions}

We shall now discuss how to express quantities, such as roots, weights, Weyl
vectors and determinants of the Cartan matrix, related to the full $n$%
-extended lattices in terms of those obtained from the reduced versions and
vice versa. We follow here largely the reasoning presented in \cite%
{gaberdiel2002class}, however, with the key difference that the node to be
removed from the full $n$-extended Dynkin diagram is not identified as the
one that decomposes the system into finite and affine diagrams, but rather
the node $\ell $ for which $D_{\ell }^{(n)}=0$. The former node might in
fact not even exist for the cases considered here. Moreover, these two types
of nodes are always different. Our construction applies to all $n$-extended
lattices.

We denote roots and weights related to the $n$-extended lattice as above by $%
\alpha _{i},\lambda _{i}$ for $i\in S=\{-n,\ldots ,0,1,\ldots ,r\}$ and
weights and roots related to the reduced system as $\tilde{\alpha}_{i},%
\tilde{\lambda}_{i}$ for $i\in \tilde{S}=S\backslash \{\ell \}=S_{1}\cup $ $%
S_{2}$. The root related to the node $\ell $ can then be expressed as%
\begin{equation}
\alpha _{\ell }=x-\nu ,~~~\ \ \text{with }\nu :=-\sum\nolimits_{i\in \tilde{S%
}}K_{\ell i}\tilde{\lambda}_{i},
\end{equation}%
where the vector $x$ is defined by the orthogonality properties $x\cdot 
\tilde{\alpha}_{i}=x\cdot \nu =0$. Consequently we have $K_{\ell \ell
}=\alpha _{\ell }^{2}=2=\nu ^{2}+x^{2}$ and the fundamental weights can be
expressed as%
\begin{equation}
\lambda _{\ell }=\frac{x}{x^{2}},~\ \ \ \ \ \lambda _{i}=\ \tilde{\lambda}%
_{i}+\left( \nu \cdot \ \tilde{\lambda}_{i}\right) \lambda _{\ell }.
\end{equation}%
Summing up the fundamental weights to construct the Weyl vector then yields
a relation between the Weyl vectors in the two respective systems%
\begin{equation}
\rho =\sum\nolimits_{i\in S}\ \lambda _{i}=\lambda _{\ell
}+\sum\nolimits_{i\in \tilde{S}}\ \lambda _{i}=\tilde{\rho}+\left( 1+\nu
\cdot \tilde{\rho}\right) \lambda _{\ell }.
\end{equation}%
Next we relate the determinants of the Cartan matrices for the two systems.
Employing Cauchy's expansion theorem for bordered matrices, see e.g. \cite%
{aitken2017}, we have%
\begin{equation}
\det K=K_{\ell \ell }\det \tilde{K}-\sum\nolimits_{i,j\in \tilde{S}}K_{\ell
i}(\limfunc{adj}\tilde{K})_{ij}K_{j\ell },  \label{cauchy}
\end{equation}%
where $\limfunc{adj}\tilde{K}$ denotes the adjugate matrix of $\tilde{K}$,
i.e. the transpose of its cofactor matrix. Recalling that $(\limfunc{adj}%
\tilde{K})_{ij}=\tilde{K}_{ij}^{-1}\det \tilde{K}$, $\tilde{K}%
_{ij}^{-1}=\lambda _{i}\cdot \lambda _{j}$ and $K_{\ell \ell }=2$, relation (%
\ref{cauchy}) can be re-expressed as%
\begin{equation}
\det K=\left( 2-\nu ^{2}\right) \det \tilde{K}.  \label{detK}
\end{equation}%
To illustrate the working of this formula and at the same time to check our
expressions from above for consistency, we present explicitly two examples
from the tables 3 and 4.

\paragraph{Example $D_{17}^{(1)}=E_{8}^{(5)}\diamond L\diamond D_{4}:$}

With $\nu =\lambda _{1}^{D_{4}}+\lambda _{-5}^{E_{8}^{(5)}}$, $\left(
\lambda _{1}^{D_{4}}\right) ^{2}=1$, $\left( \lambda
_{-5}^{E_{8}^{(5)}}\right) ^{2}=4/5$ we compute $\nu ^{2}=9/5$. Furthermore
we calculate the determinants $\det K_{D_{17}^{(1)}}=-4$, $\det
K_{E_{8}^{(5)}}=-5$, $\det K_{D_{4}}=4$ and hence confirm formula (\ref{detK}%
).

\paragraph{Example $D_{25}^{(2)}=E_{7}^{(1)}\diamond L\diamond D_{18}:$}

With $\nu =\lambda _{1}^{D_{8}}+\lambda _{-1}^{E_{7}^{(1)}}$, $\left(
\lambda _{1}^{D_{8}}\right) ^{2}=1$, $\left( \lambda
_{-1}^{E_{7}^{(1)}}\right) ^{2}=0$ we compute $\nu ^{2}=1$. We also
calculate the determinants $\det K_{D_{25}^{(2)}}=-8$, $\det
K_{E_{7}^{(1)}}=-2$, $\det K_{D_{18}}=4$ and hence confirming once more
formula (\ref{detK}).

\subsection{Decomposition of the very extended $D_{25}$-algebra aka $k_{28}$}

Let us now elaborate further on the last example. As is clear from the
above, the construction of extended Dynkin diagrams, or equivalently the
corresponding Cartan matrices, of Lorentzian Kac-Moody algebras can be
carried out in many alternative ways. As a detailed example we present now
the case of the very extended $D_{25}$-diagram, that is the $D_{25}^{(2)}$%
-algebra in our notation. It has the following Dynkin diagram:

\setlength{\unitlength}{0.58cm} 
\begin{picture}(14.00,5.0)(-1.0,4)
\thicklines

\put(-2.0,8.0){\Large{$\bullet$}}
\put(22.2,5.2){\line(1,-1){1.0}}
\put(-1.5,8.1){{$\alpha_{-2}$}}

\put(-1.0,7.0){\Large{$\bullet$}}
\put(-0.8,7.2){\line(-1,1){1.0}}
\put(-0.5,7.1){{$\alpha_{-1}$}}

\put(0.0,6.0){\Large{$\bullet$}}
\put(0.2,6.2){\line(-1,1){1.0}}
\put(0.5,6.1){{$\alpha_{0}$}}

\put(0.0,4.0){\Large{$\bullet$}}
\put(-0.8,4.0){{$\alpha_1$}}

\put(1.2,5.2){\line(-1,-1){1.0}}
\put(1.2,5.2){\line(-1,1){1.0}}
\put(1.0,5){\Large{$\bullet$}}
\put(1.2,5.2){\line(1,0){1.0}}
\put(1.0,4.5){{$\alpha_2$}}

\put(2.0,5){\Large{$\bullet$}}
\put(2.2,5.2){\line(1,0){1.0}}
\put(2.0,4.5){{$\alpha_3$}}

\put(3.0,5){\Large{$\bullet$}}
\put(3.2,5.2){\line(1,0){1.0}}
\put(3.0,4.5){{$\alpha_4$}}

\put(4.0,5){\Large{$\bullet$}}
\put(4.2,5.2){\line(1,0){0.9}}
\put(4.0,4.5){{$\alpha_5$}}

\put(5.0,5){\Large{$\circ$}}
\put(5.4,5.2){\line(1,0){0.9}}
\put(5.0,4.5){{$\alpha_6$}}

\put(6.0,5){\Large{$\bullet$}}
\put(6.2,5.2){\line(1,0){1.0}}
\put(6.0,4.5){{$\alpha_7$}}

\put(7.0,5){\Large{$\bullet$}}
\put(7.2,5.2){\line(1,0){1.0}}
\put(7.0,4.5){{$\alpha_8$}}

\put(8.0,5){\Large{$\bullet$}}
\put(8.2,5.2){\line(1,0){1.0}}
\put(8.0,4.5){{$\alpha_9$}}

\put(9.0,5){\Large{$\bullet$}}
\put(9.2,5.2){\line(1,0){1.0}}
\put(9.0,4.5){{$\alpha_{10}$}}

\put(10.0,5){\Large{$\bullet$}}
\put(10.2,5.2){\line(1,0){1.0}}
\put(10.0,4.5){{$\alpha_{11}$}}

\put(11.0,5){\Large{$\bullet$}}
\put(11.2,5.2){\line(1,0){1.0}}
\put(11.0,4.5){{$\alpha_{12}$}}

\put(12.0,5){\Large{$\bullet$}}
\put(12.2,5.2){\line(1,0){1.0}}
\put(12.0,4.5){{$\alpha_{13}$}}

\put(13.0,5){\Large{$\bullet$}}
\put(13.2,5.2){\line(1,0){1.0}}
\put(13.0,4.5){{$\alpha_{14}$}}

\put(14.0,5){\Large{$\bullet$}}
\put(14.2,5.2){\line(1,0){1.0}}
\put(14.0,4.5){{$\alpha_{15}$}}

\put(15.0,5){\Large{$\bullet$}}
\put(15.2,5.2){\line(1,0){1.0}}
\put(15.0,4.5){{$\alpha_{16}$}}

\put(16.0,5){\Large{$\bullet$}}
\put(16.2,5.2){\line(1,0){1.0}}
\put(16.0,4.5){{$\alpha_{17}$}}

\put(17.0,5){\Large{$\bullet$}}
\put(17.2,5.2){\line(1,0){1.0}}
\put(17.0,4.5){{$\alpha_{18}$}}

\put(18.0,5){\Large{$\bullet$}}
\put(18.2,5.2){\line(1,0){1.0}}
\put(18.0,4.5){{$\alpha_{19}$}}

\put(19.0,5){\Large{$\bullet$}}
\put(19.2,5.2){\line(1,0){1.0}}
\put(19.0,4.5){{$\alpha_{20}$}}

\put(20.0,5){\Large{$\bullet$}}
\put(20.2,5.2){\line(1,0){1.0}}
\put(20.0,4.5){{$\alpha_{21}$}}

\put(21.0,5){\Large{$\bullet$}}
\put(21.2,5.2){\line(1,0){1.0}}
\put(21.0,4.5){{$\alpha_{22}$}}

\put(22.0,5){\Large{$\bullet$}}
\put(22.2,5.2){\line(1,1){1.0}}
\put(22.2,5.2){\line(1,-1){1.0}}
\put(22.5,5.1){{$\alpha_{23}$}}

\put(23.0,6.0){\Large{$\bullet$}}
\put(23.0,6.6){{$\alpha_{24}$}}

\put(23.0,4.0){\Large{$\bullet$}}
\put(23.0,3.6){{$\alpha_{25}$}}
\end{picture}

\begin{center}
$D_{25}^{(2)}$-Dynkin diagram on the root lattice for $D_{25}\oplus \Pi
^{1,1}\oplus \Pi ^{1,1}$
\end{center}

The algebra belongs to the special class of hyperbolic Kac-Moody algebras
singled out by Gaberdiel, Olive and West \cite{gaberdiel2002class}, that
posses at least one node that when removed leaves a set of disconnected
Dynkin diagrams of finite type with at most one being of affine type.
Indeed, when removing the node corresponding to the root labeled by $\alpha
_{6}$, we are left with a disconnected diagram of which one corresponds to
the finite dimensional $D_{19}$-algebra and the other to the affine $%
E_{7}^{(0)}$-algebra. The corresponding root space is constructed as
indicated in (\ref{roots}).

Here, we are especially interested in the construction of the reduced Dynkin
diagram corresponding to $E_{7}^{(2)}\diamond D_{18}$. Instead of the
representation (\ref{roots}), we may also represent the roots as 
\begin{eqnarray}
\beta _{1} &:&=\alpha _{8}+\ell \text{,~~~\ \ ~~~}\beta _{i}:=\alpha _{i+7}%
\text{, }i=2,\ldots ,18, \\
\gamma _{i} &:&=\alpha _{i}\text{, }i=1,\ldots ,4\text{,~}\gamma
_{5}:=\alpha _{0}\text{, }\gamma _{6}:=\alpha _{-1}\text{, }\gamma
_{7}:=\alpha _{-2}\text{,~}\gamma _{0}:=\alpha _{5}-\bar{k}\text{, } \\
\gamma _{-1} &:&=-(k+\bar{k})-\ell \text{, }\gamma _{-2}:=-(\ell +\bar{\ell}%
).
\end{eqnarray}%
Using the standard rules for the construction of Dynkin diagrams we obtain
the same diagram as above:

\setlength{\unitlength}{0.58cm} 
\begin{picture}(14.00,5.0)(-1.0,4)
\thicklines

\put(-2.0,8.0){\Large{$\bullet$}}
\put(22.2,5.2){\line(1,-1){1.0}}
\put(-1.5,8.1){{$\gamma_{7}$}}

\put(-1.0,7.0){\Large{$\bullet$}}
\put(-0.8,7.2){\line(-1,1){1.0}}
\put(-0.5,7.1){{$\gamma_{6}$}}

\put(0.0,6.0){\Large{$\bullet$}}
\put(0.2,6.2){\line(-1,1){1.0}}
\put(0.5,6.1){{$\gamma_{5}$}}

\put(0.0,4.0){\Large{$\bullet$}}
\put(-0.8,4.0){{$\gamma_1$}}

\put(1.2,5.2){\line(-1,-1){1.0}}
\put(1.2,5.2){\line(-1,1){1.0}}
\put(1.0,5){\Large{$\bullet$}}
\put(1.2,5.2){\line(1,0){1.0}}
\put(1.0,4.5){{$\gamma_2$}}

\put(2.0,5){\Large{$\bullet$}}
\put(2.2,5.2){\line(1,0){1.0}}
\put(2.0,4.5){{$\gamma_3$}}

\put(3.0,5){\Large{$\bullet$}}
\put(3.2,5.2){\line(1,0){1.0}}
\put(3.0,4.5){{$\gamma_4$}}

\put(4.0,5){\Large{$\bullet$}}
\put(4.2,5.2){\line(1,0){0.9}}
\put(4.0,4.5){{$\gamma_0$}}

\put(5.0,5){\Large{$\bullet$}}
\put(5.3,5.2){\line(1,0){0.9}}
\put(5.0,4.5){{$\gamma_{-1}$}}

\put(6.0,5){\Large{$\bullet$}}
\put(6.2,5.2){\line(1,0){1.0}}
\put(6.0,4.5){{$\gamma_{-2}$}}

\color{red}{
\put(7.0,5){\Large{$\bullet$}}
\put(7.2,5.2){\line(1,0){1.0}}
\put(7.0,4.5){{$\beta_1$}}

\put(8.0,5){\Large{$\bullet$}}
\put(8.2,5.2){\line(1,0){1.0}}
\put(8.0,4.5){{$\beta_2$}}

\put(9.0,5){\Large{$\bullet$}}
\put(9.2,5.2){\line(1,0){1.0}}
\put(9.0,4.5){{$\beta_{3}$}}

\put(10.0,5){\Large{$\bullet$}}
\put(10.2,5.2){\line(1,0){1.0}}
\put(10.0,4.5){{$\beta_{4}$}}

\put(11.0,5){\Large{$\bullet$}}
\put(11.2,5.2){\line(1,0){1.0}}
\put(11.0,4.5){{$\beta_{5}$}}

\put(12.0,5){\Large{$\bullet$}}
\put(12.2,5.2){\line(1,0){1.0}}
\put(12.0,4.5){{$\beta_{6}$}}

\put(13.0,5){\Large{$\bullet$}}
\put(13.2,5.2){\line(1,0){1.0}}
\put(13.0,4.5){{$\beta_{7}$}}

\put(14.0,5){\Large{$\bullet$}}
\put(14.2,5.2){\line(1,0){1.0}}
\put(14.0,4.5){{$\beta_{8}$}}

\put(15.0,5){\Large{$\bullet$}}
\put(15.2,5.2){\line(1,0){1.0}}
\put(15.0,4.5){{$\beta_{9}$}}

\put(16.0,5){\Large{$\bullet$}}
\put(16.2,5.2){\line(1,0){1.0}}
\put(16.0,4.5){{$\beta_{10}$}}

\put(17.0,5){\Large{$\bullet$}}
\put(17.2,5.2){\line(1,0){1.0}}
\put(17.0,4.5){{$\beta_{11}$}}

\put(18.0,5){\Large{$\bullet$}}
\put(18.2,5.2){\line(1,0){1.0}}
\put(18.0,4.5){{$\beta_{12}$}}

\put(19.0,5){\Large{$\bullet$}}
\put(19.2,5.2){\line(1,0){1.0}}
\put(19.0,4.5){{$\beta_{13}$}}

\put(20.0,5){\Large{$\bullet$}}
\put(20.2,5.2){\line(1,0){1.0}}
\put(20.0,4.5){{$\beta_{14}$}}

\put(21.0,5){\Large{$\bullet$}}
\put(21.2,5.2){\line(1,0){1.0}}
\put(21.0,4.5){{$\beta_{15}$}}

\put(22.0,5){\Large{$\bullet$}}
\put(22.2,5.2){\line(1,1){1.0}}
\put(22.2,5.2){\line(1,-1){1.0}}
\put(22.5,5.1){{$\beta_{16}$}}

\put(23.0,6.0){\Large{$\bullet$}}
\put(22.7,6.6){{$\beta_{17}$}}

\put(23.0,4.0){\Large{$\bullet$}}
\put(22.7,3.4){{$\beta_{18}$}}}

\end{picture}

\begin{center}
$E_{7}^{(2)}\diamond D_{18}$-Dynkin diagram on the root lattice for$%
~E_{7}^{(0)}\oplus \Pi ^{1,1}\oplus \Pi ^{1,1}\oplus D_{18}$
\end{center}

The construction differs form the previous one in the sense that we have not
used the standard representation for the over extended and very extended
root, but have now linked the very extended root $\gamma _{-2}$ of $%
E_{7}^{(2)}$ with a simple root $\beta _{1}$ of the semisimple Lie algebra $%
D_{18}$. Deleting now $\bar{\ell}$ has the effect that the two links
connecting $\gamma _{-2}$ are severed so that this algebra decomposes into $%
E_{7}^{(1)}\oplus \Pi ^{1,1}\oplus D_{18}$. Thus $\gamma _{-2}=-\ell $ has
becomes a separate disconnected root of zero length $\gamma _{-2}\cdot
\gamma _{-2}=\ell ^{2}=0$. In addition, we obtain two separate disconnected
Dynkin diagrams for the over extended algebra $E_{7}^{(1)}$ and the
semisimple Lie algebra $D_{18}$:

\setlength{\unitlength}{0.58cm} 
\begin{picture}(14.00,5.0)(-1.0,4)
\thicklines

\put(-2.0,8.0){\Large{$\bullet$}}

\put(-1.5,8.1){{$\gamma_{7}$}}

\put(-1.0,7.0){\Large{$\bullet$}}
\put(-0.8,7.2){\line(-1,1){1.0}}
\put(-0.5,7.1){{$\gamma_{6}$}}

\put(0.0,6.0){\Large{$\bullet$}}
\put(0.2,6.2){\line(-1,1){1.0}}
\put(0.5,6.1){{$\gamma_{5}$}}

\put(0.0,4.0){\Large{$\bullet$}}
\put(-0.8,4.0){{$\gamma_1$}}

\put(1.2,5.2){\line(-1,-1){1.0}}
\put(1.2,5.2){\line(-1,1){1.0}}
\put(1.0,5){\Large{$\bullet$}}
\put(1.2,5.2){\line(1,0){1.0}}
\put(1.0,4.5){{$\gamma_2$}}

\put(2.0,5){\Large{$\bullet$}}
\put(2.2,5.2){\line(1,0){1.0}}
\put(2.0,4.5){{$\gamma_3$}}

\put(3.0,5){\Large{$\bullet$}}
\put(3.2,5.2){\line(1,0){1.0}}
\put(3.0,4.5){{$\gamma_4$}}

\put(4.0,5){\Large{$\bullet$}}
\put(4.2,5.2){\line(1,0){0.9}}
\put(4.0,4.5){{$\gamma_0$}}

\put(5.0,5){\Large{$\bullet$}}
\put(5.0,4.5){{$\gamma_{-1}$}}

\color{blue}{
\put(6.0,5){\Large{$\bullet$}}
\put(6.0,5.5){{$-\ell$}}}

\color{red}{
\put(7.0,5){\Large{$\bullet$}}
\put(7.2,5.2){\line(1,0){1.0}}
\put(7.0,4.5){{$\beta_1$}}

\put(8.0,5){\Large{$\bullet$}}
\put(8.2,5.2){\line(1,0){1.0}}
\put(8.0,4.5){{$\beta_2$}}

\put(9.0,5){\Large{$\bullet$}}
\put(9.2,5.2){\line(1,0){1.0}}
\put(9.0,4.5){{$\beta_{3}$}}

\put(10.0,5){\Large{$\bullet$}}
\put(10.2,5.2){\line(1,0){1.0}}
\put(10.0,4.5){{$\beta_{4}$}}

\put(11.0,5){\Large{$\bullet$}}
\put(11.2,5.2){\line(1,0){1.0}}
\put(11.0,4.5){{$\beta_{5}$}}

\put(12.0,5){\Large{$\bullet$}}
\put(12.2,5.2){\line(1,0){1.0}}
\put(12.0,4.5){{$\beta_{6}$}}

\put(13.0,5){\Large{$\bullet$}}
\put(13.2,5.2){\line(1,0){1.0}}
\put(13.0,4.5){{$\beta_{7}$}}

\put(14.0,5){\Large{$\bullet$}}
\put(14.2,5.2){\line(1,0){1.0}}
\put(14.0,4.5){{$\beta_{8}$}}

\put(15.0,5){\Large{$\bullet$}}
\put(15.2,5.2){\line(1,0){1.0}}
\put(15.0,4.5){{$\beta_{9}$}}

\put(16.0,5){\Large{$\bullet$}}
\put(16.2,5.2){\line(1,0){1.0}}
\put(16.0,4.5){{$\beta_{10}$}}

\put(17.0,5){\Large{$\bullet$}}
\put(17.2,5.2){\line(1,0){1.0}}
\put(17.0,4.5){{$\beta_{11}$}}

\put(18.0,5){\Large{$\bullet$}}
\put(18.2,5.2){\line(1,0){1.0}}
\put(18.0,4.5){{$\beta_{12}$}}

\put(19.0,5){\Large{$\bullet$}}
\put(19.2,5.2){\line(1,0){1.0}}
\put(19.0,4.5){{$\beta_{13}$}}

\put(20.0,5){\Large{$\bullet$}}
\put(20.2,5.2){\line(1,0){1.0}}
\put(20.0,4.5){{$\beta_{14}$}}

\put(21.0,5){\Large{$\bullet$}}
\put(21.2,5.2){\line(1,0){1.0}}
\put(21.0,4.5){{$\beta_{15}$}}

\put(22.0,5){\Large{$\bullet$}}
\put(22.2,5.2){\line(1,1){1.0}}
\put(22.2,5.2){\line(1,-1){1.0}}
\put(22.5,5.1){{$\beta_{16}$}}

\put(23.0,6.0){\Large{$\bullet$}}
\put(22.7,6.6){{$\beta_{17}$}}

\put(23.0,4.0){\Large{$\bullet$}}
\put(22.7,3.4){{$\beta_{18}$}}}

\end{picture}

\begin{center}
Reduced Dynkin diagram of $D_{25}^{(2)}=E_{7}^{(1)}\diamond L\diamond D_{18}$
\end{center}

We also notice that the root $\alpha _{\ell }=\alpha _{7}$ for which $%
D_{\ell }=0$ is different from the root $\alpha _{6}$ that need to be chosen
for the very extended root lattice to reduce to an affine and a finite Kac
Moody algebra.

\subsection{Examples for double and triple decompositions}

As indicated in the tables 3 and 4 above, there exist also $n$-extended
algebras for which there are two or even three nodes, say $\ell ,\ell
^{\prime }$ and $\ell ^{\prime \prime }$ , for which $D_{\ell }=D_{\ell
^{\prime }}=D_{\ell ^{\prime \prime }}=0$. We present here two examples of
Dynkin diagrams that decompose on the semisimple part as well as on the
extended part. For instance, we have a triple decomposition for%
\begin{equation*}
\begin{tabular}{lllll}
&  & $E_{6}^{(8)}\diamond L^{2}\diamond A_{18}$ &  &  \\ 
& $\nearrow $ &  & $\searrow $ &  \\ 
$A_{24}^{(10)}$ &  &  &  & $E_{6}^{(3)}\diamond L\diamond A_{4}\diamond
L^{2}\diamond A_{18}$ \\ 
& $\searrow $ &  & $\nearrow $ &  \\ 
&  & $A_{24}^{(5)}\diamond L\diamond A_{4}$ &  & 
\end{tabular}%
\end{equation*}%
for which the final disconnected Dynkin diagram is :

\begin{picture}(10.00,15.0)(0.0,4.0)
\thicklines
{\color{blue}{
\put(0.1,5.2){\line(4,1){11.5}}
\put(23.3,5.2){\line(-4,1){11.5}}		
\put(0.0,5){\Large{$\bullet$}}
\put(0.2,5.2){\line(1,0){1.0}}
\put(1.0,5){\Large{$\bullet$}}}}
\put(0.0,4.5){{$\alpha_1$}}
\put(1.0,4.5){{$\alpha_2$}}
{\color{red}{
\put(2.0,5){\Large{$\bullet$}}}}
\put(2.0,4.5){{$\alpha_3$}}
\put(3.0,5){\Large{$\bullet$}}
\put(3.2,5.2){\line(1,0){1.0}}
\put(3.0,4.5){{$\alpha_4$}}
\put(4.0,5){\Large{$\bullet$}}
\put(4.2,5.2){\line(1,0){0.9}}
\put(4.0,4.5){{$\alpha_5$}}
\put(5.0,5){\Large{$\bullet$}}
\put(5.3,5.2){\line(1,0){0.9}}
\put(5.0,4.5){{$\alpha_6$}}
\put(6.0,5){\Large{$\bullet$}}
\put(6.2,5.2){\line(1,0){1.0}}
\put(6.0,4.5){{$\alpha_7$}}
\put(7.0,5){\Large{$\bullet$}}
\put(7.2,5.2){\line(1,0){1.0}}
\put(7.0,4.5){{$\alpha_8$}}
\put(8.0,5){\Large{$\bullet$}}
\put(8.2,5.2){\line(1,0){1.0}}
\put(8.0,4.5){{$\alpha_9$}}
\put(9.0,5){\Large{$\bullet$}}
\put(9.2,5.2){\line(1,0){1.0}}
\put(9.0,4.5){{$\alpha_{10}$}}
\put(10.0,5){\Large{$\bullet$}}
\put(10.2,5.2){\line(1,0){1.0}}
\put(10.0,4.5){{$\alpha_{11}$}}
\put(11.0,5){\Large{$\bullet$}}
\put(11.2,5.2){\line(1,0){1.0}}
\put(11.0,4.5){{$\alpha_{12}$}}
\put(12.0,5){\Large{$\bullet$}}
\put(12.2,5.2){\line(1,0){1.0}}
\put(12.0,4.5){{$\alpha_{13}$}}
\put(13.0,5){\Large{$\bullet$}}
\put(13.2,5.2){\line(1,0){1.0}}
\put(13.0,4.5){{$\alpha_{14}$}}
\put(14.0,5){\Large{$\bullet$}}
\put(14.2,5.2){\line(1,0){1.0}}
\put(14.0,4.5){{$\alpha_{15}$}}
\put(15.0,5){\Large{$\bullet$}}
\put(15.2,5.2){\line(1,0){1.0}}
\put(15.0,4.5){{$\alpha_{16}$}}
\put(16.0,5){\Large{$\bullet$}}
\put(16.2,5.2){\line(1,0){1.0}}
\put(16.0,4.5){{$\alpha_{17}$}}
\put(17.0,5){\Large{$\bullet$}}
\put(17.2,5.2){\line(1,0){1.0}}
\put(17.0,4.5){{$\alpha_{18}$}}
\put(18.0,5){\Large{$\bullet$}}
\put(18.2,5.2){\line(1,0){1.0}}
\put(18.0,4.5){{$\alpha_{19}$}}
\put(19.0,5){\Large{$\bullet$}}
\put(19.2,5.2){\line(1,0){1.0}}
\put(19.0,4.5){{$\alpha_{20}$}}
\put(20.0,5){\Large{$\bullet$}}
\put(20.0,4.5){{$\alpha_{21}$}}
{\color{red}{
\put(10.6,13.9){L}
\put(11.5,13.9){\Large{$\bullet$}}
\put(21.0,5){\Large{$\bullet$}}}}

\put(11.5,14.9){\Large{$\bullet$}}
\put(11.5,15.9){\Large{$\bullet$}}
\put(11.5,16.9){\Large{$\bullet$}}
\put(11.5,17.9){\Large{$\bullet$}}
\put(11.7,15.2){\line(0,1){1.0}}
\put(11.7,16.2){\line(0,1){1.0}}
\put(11.7,17.2){\line(0,1){1.0}}
\put(21.0,4.5){{$\alpha_{22}$}}

\put(22.0,4.5){{$\alpha_{23}$}}

\put(23.0,4.5){{$\alpha_{24}$}}
{\color{blue}{
\put(22.0,5){\Large{$\bullet$}}
\put(22.2,5.2){\line(1,0){1.0}}
\put(23.0,5){\Large{$\bullet$}}
\put(11.5,7.9){\Large{$\bullet$}}
\put(11.7,8.2){\line(0,1){1.0}}
\put(11.5,8.9){\Large{$\bullet$}}
\put(11.7,9.2){\line(0,1){1.0}}
\put(11.5,9.9){\Large{$\bullet$}}
\put(11.7,10.2){\line(0,1){1.0}}
\put(11.5,10.9){\Large{$\bullet$}}
\put(11.7,11.2){\line(0,1){1.0}}
\put(11.5,11.9){\Large{$\bullet$}}
\put(11.7,12.2){\line(0,1){1.0}}
\put(10.4,10.4){$E_6^{(3)}$}
\put(11.5,12.9){\Large{$\bullet$}}}}
\put(10.4,16.4){$A_4$}

\put(12.4, 8.1){{$\alpha_{0}$}}
\put(12.4, 9.0){{$\alpha_{-1}$}}
\put(12.4, 10.0){{$\alpha_{-2}$}}
\put(12.4, 11.0){{$\alpha_{-3}$}}
\put(12.4, 12.0){{$\alpha_{-4}$}}
\put(12.4, 13.0){{$\alpha_{-5}$}}
\put(12.4, 14.0){{$\alpha_{-6}$}}
\put(12.4, 15.0){{$\alpha_{-7}$}}
\put(12.4, 16.0){{$\alpha_{-8}$}}
\put(12.4, 17.0){{$\alpha_{-9}$}}
\put(12.4, 18.0){{$\alpha_{-10}$}}
\end{picture}

\begin{center}
Reduced Dynkin diagram of $A_{24}^{(10)} = E_{6}^{(3)}\diamond L\diamond
A_{4}\diamond L^{2}\diamond A_{18}$
\end{center}

\noindent Similarly, $D_{36}^{(14)}$ doubly decomposes as 
\begin{equation*}
\begin{tabular}{lllll}
&  & $D_{36}^{(10)}\diamond L\diamond A_{3}$ &  &  \\ 
& $\nearrow $ &  & $\searrow $ &  \\ 
$D_{36}^{(14)}$ &  &  &  & $A_{3}\diamond L\diamond E_{6}^{(6)}\diamond
L\diamond D_{31}$ \\ 
& $\searrow $ &  & $\nearrow $ &  \\ 
&  & $E_{8}^{(10)}\diamond L\diamond D_{31}$ &  & 
\end{tabular}%
\end{equation*}%
Further examples can be obtained from tables 3 and 4 for the cases with bold
entries.

\section{Roots, weights, Weyl vectors and decomposition of the $\hat{A}%
_{r}^{(n,m)}$-algebras}

Since the $\hat{A}_{r}^{(n,m)}$-algebras occur naturally in the
decomposition of the $n$-extended Lorentzian Kac-Moody algebras we shall now
discuss them in further detail, with particular emphasis on their
decomposition. The corresponding Dynkin diagrams are equivalent to those
arising in the description of the so-called $T_{p,q,r}$-singularities\ \cite%
{katz1997mirror}, with the identification $\hat{A}_{p+q+1}^{(r,p+1)}\equiv
T_{p,q,r}$. We represent the simple $\hat{A}_{r}^{(n,m)}$-roots in terms of
the $r$ simple roots $\alpha _{i}~$of the semisimple Lie algebra $\mathbf{g}$
and the Lorentzian roots, with the $m^{\text{th}}$ root modified similarly
as the affine root for the $n$-extended algebras $\alpha _{m}\rightarrow
\alpha _{m}+k_{1}$. Thus the $r+n$ simple $\hat{A}_{r}^{(n,m)}$-roots are
represented as 
\begin{equation}
\hat{\alpha}=\left\{ \alpha _{1},\ldots ,\alpha _{m-1},\alpha
_{m}+k_{1},\alpha _{m+1},\ldots ,\alpha _{r},\alpha _{-1}=-k_{1}-\bar{k}%
_{1},\ldots ,\alpha _{-j}=k_{j-1}-k_{j}-\bar{k}_{j}\right\} ,
\end{equation}%
with $~j=2,\ldots ,n$. Using the orthogonality relation%
\begin{equation}
\lambda _{i}^{(n,m)}\cdot \hat{\alpha}_{j}=\delta _{ij},\qquad i,j=-n,\ldots
,-1,1,\ldots ,r,
\end{equation}%
together with $\lambda _{i}^{(n,m)}=\sum_{j=1}^{n+r}\hat{K}_{ij}^{-1}\hat{%
\alpha}_{j}$, $\hat{K}_{ij}^{-1}=\lambda _{i}^{(n,m)}\cdot \lambda
_{i}^{(n,m)}$, we can construct the $n+r$ fundamental weights. We shall
focus here on the case for which the extension is attached onto the middle
node $\hat{A}_{r=2\ell +1}^{(n,\ell +1)}$, so that $m=h/2$, and refer to
them as $\hat{A}_{r}^{(n)}$. We find in this case the fundamental weights 
\begin{eqnarray}
\hat{\lambda}_{i}^{(n)} &=&\lambda _{i}^{f}+\frac{2n}{nh-4(n+1)}\min
(i,h-i)\left( \lambda _{0}^{(n)}-\lambda _{h/2}^{f}\right) ,~~\ \ \
i=1,\ldots ,r, \\
\hat{\lambda}_{-j}^{(n)} &=&\lambda _{-j}^{(n)}+\frac{4(n-j+1)}{nh-4(n+1)}%
\left( \lambda _{0}^{(n)}-\lambda _{h/2}^{f}\right) ,~~~\ \ \ ~~\ \
j=1,\ldots ,n,
\end{eqnarray}%
where $\lambda _{i}^{f}~$\ are the fundamental weights of $A_{r}$ and $%
\lambda _{0}^{(n)}$, $\lambda _{-j}^{(n)}$ are fundamental weights for the $n
$-extended Lorentzian Kac-Moody algebras as determined above in equations (%
\ref{l0}), (\ref{ln}). The Weyl vector results therefore \ to%
\begin{equation}
\hat{\rho}^{(n)}=\sum_{j=-n,j\neq 0}^{r}\hat{\lambda}_{i}^{(n)}=\rho
^{(n)}-h\lambda _{0}^{(n)}+\frac{n(h^{2}+4n+4)}{2n(h-4)-8}\left( \lambda
_{0}^{(n)}-\lambda _{m}^{f}\right) .
\end{equation}%
Next we compute the constants 
\begin{eqnarray}
\hat{D}_{i}^{(n)} &=&\hat{\rho}^{(n)}\cdot \hat{\lambda}_{i}^{(n)}=\frac{%
n(4+4n+h^{2})}{16+4n(4-h)}\min (i,h-i)+\frac{i}{2}(h-i),~~\ \ \ i=1,\ldots
,r, \\
\hat{D}_{-j}^{(n)} &=&\hat{\rho}^{(n)}\cdot \hat{\lambda}_{-j}^{(n)}=\frac{%
(j-n-1)[h^{2}+4j(1+n)+nh(1-j)]}{2n(h-4)-8},~~~\ \ \ ~\ j=1,\ldots ,n.~~~~
\end{eqnarray}%
The algebras decompose, for the same reasons as previously argued for the $n$%
-extended algebras ,when the constants $\hat{D}^{(n)}$ vanish. We determine%
\begin{eqnarray}
\hat{D}_{i}^{(n)} &=&0,~~\ ~\text{for \ }i=\frac{n(4n+4+h^{2})}{2n(h-4)-8},h-%
\frac{n(4n+4+h^{2})}{2n(h-4)-8},  \label{Dhi} \\
\hat{D}_{-j}^{(n)} &=&0,~~\ ~\text{for \ }j=\frac{h(h+n)}{n(h-4)-4},
\label{Dhj}
\end{eqnarray}%
Thus the only meaningful solutions, i.e. those for which $i,\in \mathbb{N}$, 
$i\leq r$, to (\ref{Dhi}) give rise to the decompostions on the leg of the
Dynkin diagram corresponding to the $A_{r}$-diagram as listed in table 5.

\begin{center}
\begin{tabular}{|l|l|l|}
\hline
$\hat{A}_{13}^{(2)}=\hat{A}_{11}^{(2)}\diamond L^{2}$ & $\hat{A}_{13}^{(4)}=%
\hat{A}_{9}^{(4)}\diamond L^{2}\diamond A_{1}^{2}$ & $\hat{A}_{13}^{(6)}=%
\mathbf{\hat{A}}_{\mathbf{9}}^{\mathbf{(6)}}\diamond L^{2}\diamond A_{1}^{2}$
\\ \hline
$\hat{A}_{13}^{(13)}=\mathbf{\hat{A}}_{\mathbf{11}}^{\mathbf{(13)}}\diamond
L^{2}$ & $\hat{A}_{17}^{(2)}=\hat{A}_{9}^{(2)}\diamond L^{2}\diamond
A_{3}^{2}$ & $\hat{A}_{17}^{(14)}=\mathbf{\hat{A}}_{\mathbf{9}}^{\mathbf{(14)%
}}\diamond L^{2}\diamond A_{3}^{2}$ \\ \hline
$\hat{A}_{19}^{(1)}=\hat{A}_{13}^{(1)}\diamond L^{2}\diamond A_{2}^{2}$ & $%
\hat{A}_{19}^{(4)}=\hat{A}_{7}^{(4)}\diamond L^{2}\diamond A_{5}^{2}$ & $%
\hat{A}_{19}^{(7)}=\mathbf{\hat{A}}_{\mathbf{7}}^{\mathbf{(7)}}\diamond
L^{2}\diamond A_{5}^{2}$ \\ \hline
$\hat{A}_{19}^{(34)}=\mathbf{\hat{A}}_{\mathbf{13}}^{\mathbf{(34)}}\diamond
L^{2}\diamond A_{2}^{2}$ & $\hat{A}_{21}^{(3)}=\hat{A}_{7}^{(3)}\diamond
L^{2}\diamond A_{6}^{2}$ & $\hat{A}_{21}^{(10)}=\mathbf{\hat{A}}_{\mathbf{7}%
}^{\mathbf{(10)}}\diamond L^{2}\diamond A_{6}^{2}$ \\ \hline
$\hat{A}_{25}^{(1)}=\hat{A}_{11}^{(1)}\diamond L^{2}\diamond A_{6}^{2}$ & $%
\hat{A}_{25}^{(38)}=\mathbf{\hat{A}}_{\mathbf{11}}^{\mathbf{(38)}}\diamond
L^{2}\diamond A_{6}^{2}$ & $\hat{A}_{29}^{(2)}=\hat{A}_{7}^{(2)}\diamond
L^{2}\diamond A_{10}^{2}$ \\ \hline
$\hat{A}_{29}^{(19)}=\mathbf{\hat{A}}_{\mathbf{7}}^{\mathbf{(19)}}\diamond
L^{2}\diamond A_{10}^{2}$ & $\hat{A}_{41}^{(6)}=E_{6}^{(4)}\diamond
L^{2}\diamond A_{17}^{2}$ & $\hat{A}_{41}^{(8)}=\mathbf{E}_{\mathbf{6}}^{%
\mathbf{(6)}}\diamond L^{2}\diamond A_{17}^{2}$ \\ \hline
$\hat{A}_{43}^{(1)}=\hat{A}_{9}^{(1)}\diamond L^{2}\diamond A_{16}^{2}$ & $%
\hat{A}_{43}^{(5)}=E_{6}^{(3)}\diamond L^{2}\diamond A_{18}^{2}$ & $\hat{A}%
_{43}^{(54)}=\hat{A}_{7}^{(54)}\diamond L^{2}\diamond A_{16}^{2}$ \\ \hline
$\hat{A}_{43}^{(10)}=\mathbf{E}_{\mathbf{6}}^{\mathbf{(8)}}\diamond
L^{2}\diamond A_{18}^{2}$ & $\hat{A}_{49}^{(4)}=E_{6}^{(2)}\diamond
L^{2}\diamond A_{21}^{2}$ & $\hat{A}_{49}^{(14)}=\mathbf{E}_{\mathbf{6}}^{%
\mathbf{(12)}}\diamond L^{2}\diamond A_{21}^{2}$ \\ \hline
$\hat{A}_{71}^{(3)}=E_{6}^{(1)}\diamond L^{2}\diamond A_{32}^{2}$ & $\hat{A}%
_{71}^{(26)}=\mathbf{E}_{\mathbf{6}}^{\mathbf{(24)}}\diamond L^{2}\diamond
A_{32}^{2}$ &  \\ \hline
\end{tabular}

\bigskip Table 5: Decomposition of the algebras $A_{r}^{(n)}$ on the $A_{r}$%
-leg of the Dynkin diagram.
\end{center}

\noindent On the extended leg of the Dynkin diagram we find with $j\in 
\mathbb{N}$, $j\leq n$, the solutions to (\ref{Dhi}) as reported in table 6.

\begin{center}
\begin{tabular}{|l|l|l|}
\hline
$\hat{A}_{7}^{(7)}=\hat{A}_{7}^{(4)}\diamond L\diamond A_{2}$ & $\hat{A}%
_{7}^{(10)}=\hat{A}_{7}^{(3)}\diamond L\diamond A_{6}$ & $\hat{A}_{7}^{(19)}=%
\hat{A}_{7}^{(2)}\diamond L\diamond A_{16}$ \\ \hline
$\hat{A}_{9}^{(6)}=\hat{A}_{9}^{(4)}\diamond L\diamond A_{1}$ & $\hat{A}%
_{9}^{(14)}=\hat{A}_{9}^{(2)}\diamond L\diamond A_{11}$ & $\hat{A}%
_{9}^{(54)}=\hat{A}_{9}^{(1)}\diamond L\diamond A_{52}$ \\ \hline
$\hat{A}_{11}^{(8)}=\hat{A}_{11}^{(3)}\diamond L\diamond A_{3}$ & $\hat{A}%
_{11}^{(13)}=\hat{A}_{11}^{(2)}\diamond L\diamond A_{10}$ & $\hat{A}%
_{11}^{(38)}=\hat{A}_{11}^{(1)}\diamond L\diamond A_{36}$ \\ \hline
$\hat{A}_{13}^{(6)}=\hat{A}_{13}^{(4)}\diamond L\diamond A_{1}$ & $\hat{A}%
_{13}^{(13)}=\hat{A}_{13}^{(2)}\diamond L\diamond A_{10}$ & $\hat{A}%
_{13}^{(34)}=\hat{A}_{13}^{(1)}\diamond L\diamond A_{32}$ \\ \hline
$\hat{A}_{15}^{(33)}=\hat{A}_{15}^{(1)}\diamond L\diamond A_{31}$ & $\hat{A}%
_{17}^{(14)}=\hat{A}_{17}^{(2)}\diamond L\diamond A_{11}$ & $\hat{A}%
_{19}^{(7)}=\hat{A}_{19}^{(4)}\diamond L\diamond A_{2}$ \\ \hline
$\hat{A}_{19}^{(34)}=\hat{A}_{19}^{(4)}\diamond L\diamond A_{2}$ & $\hat{A}%
_{21}^{(10)}=\hat{A}_{21}^{(3)}\diamond L\diamond A_{6}$ & $\hat{A}%
_{25}^{(38)}=\hat{A}_{25}^{(1)}\diamond L\diamond A_{36}$ \\ \hline
$\hat{A}_{29}^{(19)}=\hat{A}_{29}^{(2)}\diamond L\diamond A_{16}$ & $\hat{A}%
_{31}^{(13)}=\hat{A}_{31}^{(3)}\diamond L\diamond A_{9}$ & $\hat{A}%
_{31}^{(43)}=\hat{A}_{31}^{(1)}\diamond L\diamond A_{41}$ \\ \hline
$\hat{A}_{41}^{(8)}=\hat{A}_{41}^{(6)}\diamond L\diamond A_{1}$ & $\hat{A}%
_{43}^{(10)}=\hat{A}_{43}^{(5)}\diamond L\diamond A_{4}$ & $\hat{A}%
_{43}^{(54)}=\hat{A}_{43}^{(1)}\diamond L\diamond A_{52}$ \\ \hline
$\hat{A}_{49}^{(14)}=\hat{A}_{49}^{(4)}\diamond L\diamond A_{9}$ & $\hat{A}%
_{71}^{(26)}=\hat{A}_{71}^{(3)}\diamond L\diamond A_{22}$ & $\hat{A}%
_{79}^{(89)}=\hat{A}_{79}^{(1)}\diamond L\diamond A_{87}$ \\ \hline
$\hat{A}_{111}^{(13)}=\hat{A}_{111}^{(9)}\diamond L\diamond A_{3}$ & $\hat{A}%
_{127}^{(19)}=\hat{A}_{127}^{(7)}\diamond L\diamond A_{11}$ & $\hat{A}%
_{239}^{(49)}=\hat{A}_{239}^{(5)}\diamond L\diamond A_{43}$ \\ \hline
\end{tabular}

\bigskip Table 6: Decomposition of the algebras $A_{r}^{(n)}$ on the
extended leg of the Dynkin diagram.
\end{center}

Finally we consider the $\hat{A}_{r}^{(n,m)}$-algebras in general. We will
not present here a full discussion of the weight lattices, the Weyl vectors
and the constants $\hat{D}_{i}^{(n,m)}$ as for the special $\hat{A}%
_{r}^{(n)} $-case, but only list the decompositions of those cases that
appear in table 4. Our results are reported in table 7.%
\begin{equation*}
\begin{tabular}{|l|l|l|}
\hline
$\hat{A}_{14}^{(1,5)}=\hat{A}_{12}^{(1,5)}\diamond L\diamond A_{1}$ & $\hat{A%
}_{16}^{(1,7)}=\hat{A}_{13}^{(1)}\diamond L\diamond A_{2}$ & $\hat{A}%
_{19}^{(1,8)}=\hat{A}_{13}^{(1,6)}\diamond L\diamond A_{5}$ \\ \hline
$\hat{A}_{18}^{(1,6)}=\hat{A}_{11}^{(1)}\diamond L\diamond A_{6}$ & $\hat{A}%
_{19}^{(1,5)}=\hat{A}_{10}^{(1,5)}\diamond L\diamond A_{8}$ & $\hat{A}%
_{23}^{(1,6)}=\hat{A}_{10}^{(1,5)}\diamond L\diamond A_{12}$ \\ \hline
$\hat{A}_{26}^{(1,5)}=\hat{A}_{9}^{(1)}\diamond L\diamond A_{16}$ & $\hat{A}%
_{33}^{(1,8)}=E_{7}^{(4)}\diamond L\diamond A_{21}$ & $\hat{A}%
_{34}^{(1,7)}=E_{7}^{(3)}\diamond L\diamond A_{23}$ \\ \hline
$\hat{A}_{54}^{(1,5)}=E_{7}^{(1)}\diamond L\diamond A_{45}$ & $\hat{A}%
_{76}^{(1,11)}=E_{8}^{(5)}\diamond L\diamond A_{62}$ & $\hat{A}%
_{79}^{(1,10)}=E_{8}^{(4)}\diamond L\diamond A_{66}$ \\ \hline
$\hat{A}_{86}^{(1,9)}=E_{8}^{(3)}\diamond L\diamond A_{74}$ & $\hat{A}%
_{160}^{(1,7)}=E_{8}^{(2)}\diamond L\diamond A_{150}$ &  \\ \hline
\end{tabular}%
\end{equation*}

\begin{center}
Table 7: Decomposition of the algebras $\hat{A}_{r}^{(n,m)}$ that occur in
table 4.
\end{center}

$\hat{A}_{r}^{(n,m)}$-algebras that appear in table 4 and are not reported
in table 7 do not decompose. Thus, similarly as discussed in section 6.3, we
also obtain double decompositions involving these type of algebras. For
instance, we have 
\begin{equation*}
\begin{tabular}{lllll}
&  & $\hat{A}_{34}^{(1,7)}\diamond L\diamond D_{5}$ &  &  \\ 
& $\nearrow $ &  & $\searrow $ &  \\ 
$D_{14}^{(26)}$ &  &  &  & $A_{23}\diamond L\diamond E_{7}^{(3)}\diamond
L\diamond D_{5}$ \\ 
& $\searrow $ &  & $\nearrow $ &  \\ 
&  & $D_{14}^{(2)}\diamond L\diamond A_{23}$ &  & 
\end{tabular}%
\end{equation*}%
as seen from table 4, 7 and (\ref{dx}).

\section{Conclusions}

We defined and investigated a new class of Kac-Moody algebras, referred to
as $n$-extended Lorentzian Kac-Moody algebras $\mathbf{g}_{-n}$. For the
corresponding Dynkin diagrams we constructed the associated root and weight
lattices with generic expressions for all simple roots $\alpha _{i}^{(n)}$
and fundamental weights $\lambda _{i}^{(n)}$. The latter were used to derive
closed expressions for the Weyl vectors $\rho ^{(n)}$ for any value of $n$.
The signatures of the product $\rho ^{(n)}\cdot \rho ^{(n)}$, that is the
generalisation of the Freudenthal-de Vries strange formula, led to a
necessary condition for the $n$-extended Lorentzian Kac-Moody algebras to
possess a $SO(1,2)$-principal subalgebra. From the inner products of the
Weyl vector $\rho ^{(n)}$ and the fundamental weights $\lambda _{i}^{(n)}$
we compute the expansion coefficients $D_{i}^{(n)}$ for the $J_{3}$
-generator of the principal $SO(1,2)$ or $SO(3)$ subalgebra. When these
constants vanish the decomposition the corresponding Dynkin diagram can be
reduced. For the reduced diagrams we analyse in detail whether $D_{i}>0$\ or 
$\hat{D}_{i}<0$\ for all $i$, which constitutes a necessary and sufficient
condition for the existence of a $SO(3)$-principal subalgebra or a $SO(1,2)$%
-principal subalgebra, respectively. We derive explicit formulae augmented
by examples that allow to express quantities related to the $n$-extended
systems in terms of the reduced counterparts and vice versa. We provide
complete lists for \emph{all} decompositions of the $n$-extended Lorentzian
Kac-Moody algebras $\mathbf{g}_{-n}$. A similarly detailed analysis is
presented for the $A_{r}^{(n)}$-algebras, but for $\hat{A}_{r}^{(n,m)}\neq
A_{r}^{(n)}$ we only report the decomposition for the cases appearing in the
decomposition of $\mathbf{g}_{-n}$.

Besides the aforementioned applications in string theory, one may also apply
the constructions here in the context of classical and quantum integrable
systems that are formulated in terms of roots, weights or even directly in
terms of principle subalgebras, such as Toda theories \cite{Mass2},
Calogero-Moser-Sutherlund systems etc. Even though it was found that for
some of the Toda theories based on Lorentzian root systems do not pass the
Painlev\'{e} test \cite{gebert1996toda}, and are therefore not integrable,
the constructions presented here suggest that they contain some integrable
components and hence are candidates for a systematic study of nonintegrable
quantum field theories \cite{delfino1996non}.

\medskip

\noindent \textbf{Acknowledgments:} SW is supported by a City, University of
London Research Fellowship.

\newif\ifabfull\abfulltrue


\begin{thebibliography}{10}

\bibitem{kacinfinite}
V.~G. Kac,
\newblock Infinite dimensional Lie algebras,
\newblock Cambridge University Press, Cambridge  (1990).

\bibitem{goddard1985algebras}
P.~Goddard and D.~Olive,
\newblock Algebras, lattices and strings,
\newblock in {\em Vertex operators in mathematics and physics}, pages 51--96,
  Springer, 1985.

\bibitem{goddard1986kac}
P.~Goddard and D.~Olive,
\newblock Kac-Moody and Virasoro algebras in relation to quantum physics,
\newblock Int. J. of Mod. Phys. A {\bf 1}(02), 303--414 (1986).

\bibitem{fring2018lectures}
ed.~A. Fring and N.~Turok,
\newblock Lectures of David Olive on Gauge Theories and Lie Algebras,
\newblock World Scientific, Singapore, (2020).

\bibitem{witten1995string}
E.~Witten,
\newblock String theory dynamics in various dimensions,
\newblock Nucl. Phys. B {\bf 443}(1-2), 85--126 (1995).

\bibitem{west2000hidden}
P.~West,
\newblock Hidden superconformal symmetry in M-theory,
\newblock JHEP {\bf 2000}(08), 007 (2000).

\bibitem{west2001e11}
P.~West,
\newblock E11 and M-theory,
\newblock Classical and Quantum Gravity {\bf 18}(21), 4443 (2001).

\bibitem{damour200210}
T.~Damour, M.~Henneaux, and H.~Nicolai,
\newblock E10 and a small tension expansion of M theory,
\newblock Phys. Rev. Lett. {\bf 89}(22), 221601 (2002).

\bibitem{riccioni2007e11}
F.~Riccioni and P.~C. West,
\newblock The E11 origin of all maximal supergravities,
\newblock JHEP {\bf 2007}(07), 063 (2007).

\bibitem{englert2003s}
F.~Englert, L.~Houart, A.~Taormina, and P.~West,
\newblock The symmetry of M-theories,
\newblock JHEP {\bf 2003}(09), 020 (2003).

\bibitem{bergshoeff2007e11}
E.~A. Bergshoeff, T.~A. Nutma, and I.~De~Baetselier,
\newblock E11 and the embedding tensor,
\newblock JHEP {\bf 2007}(09), 047 (2007).

\bibitem{de2001hidden}
B.~De~Wit and H.~Nicolai,
\newblock Hidden symmetries, central charges and all that,
\newblock Classical and Quantum Gravity {\bf 18}(16), 3095 (2001).

\bibitem{berman2011}
D.~S. Berman and M.~J. Perry,
\newblock Generalized Geometry and M theory,
\newblock JHEP {\bf 2011}(6), 74 (2011).

\bibitem{de2008gauged}
B.~De~Wit, H.~Nicolai, and H.~Samtleben,
\newblock Gauged supergravities, tensor hierarchies, and M-theory,
\newblock JHEP {\bf 2008}(02), 044 (2008).

\bibitem{hohm2014}
O.~Hohm and H.~Samtleben,
\newblock Exceptional field theory. I. E 6 (6)-covariant form of M-theory and
  type IIB,
\newblock Phys. Rev. D {\bf 89}(6), 066016 (2014).

\bibitem{Hum}
J.~E. Humphreys,
\newblock Introduction to Lie Algebras and Representation Theory,
\newblock Springer, Berlin  (1972).

\bibitem{carbone2010}
L.~Carbone, S.~Chung, L.~Cobbs, R.~McRae, D.~Nandi, Y.~Naqvi, and D.~Penta,
\newblock Classification of hyperbolic Dynkin diagrams, root lengths and Weyl
  group orbits,
\newblock J. of Phys. A: Math. and Theor. {\bf 43}(15), 155209 (2010).

\bibitem{gebert199510}
R.~W. Gebert and H.~Nicolai,
\newblock E10 for beginners,
\newblock in {\em Strings and Symmetries}, pages 197--210, Springer, 1995.

\bibitem{nicolai2004low}
H.~Nicolai and T.~Fischbacher,
\newblock Low level representations for E10 and E11,
\newblock Cont. Math {\bf 343}, 191 (2004).

\bibitem{borcherds1988generalized}
R.~Borcherds,
\newblock Generalized Kac-Moody algebras,
\newblock Journal of Algebra {\bf 115}(2), 501--512 (1988).

\bibitem{gaberdiel2002class}
M.~R. Gaberdiel, D.~I. Olive, and P.~C. West,
\newblock A class of Lorentzian Kac--Moody algebras,
\newblock Nuclear Physics B {\bf 645}(3), 403--437 (2002).

\bibitem{brown2004m}
J.~Brown, O.~J. Ganor, and C.~Helfgott,
\newblock M-theory and E10: billiards, branes, and imaginary roots,
\newblock JHEP {\bf 2004}(08), 063 (2004).

\bibitem{henneaux2008}
M.~Henneaux, D.~Persson, and P.~Spindel,
\newblock Spacelike singularities and hidden symmetries of gravity,
\newblock Living Reviews in Relativity {\bf 11}(1), 1 (2008).

\bibitem{nicolai2001dio}
H.~Nicolai and D.~I. Olive,
\newblock The principal SO (1, 2) subalgebra of a hyperbolic Kac--Moody
  algebra,
\newblock Lett. in Math. Phys. {\bf 58}(2), 141--152 (2001).

\bibitem{burns2000}
J.~M. Burns,
\newblock An elementary proof of the strange formula of Freudenthal and de
  Vries,
\newblock Quarterly J. of Math. {\bf 51}(3), 295--297 (2000).

\bibitem{Mass2}
A.~Fring, H.~C. Liao, and D.~Olive,
\newblock The mass spectrum and coupling in affine Toda theories,
\newblock Phys. Lett. {\bf B266}, 82--86 (1991).

\bibitem{kneipp1993}
M.~A.~C. Kneipp and D.~I. Olive,
\newblock Crossing and antisolitons in affine Toda theories,
\newblock Nucl. Phys. B {\bf 408}(3), 565--578 (1993).

\bibitem{Kost}
B.~Kostant,
\newblock The principal three-dimensional subgroup and the Betti numbers of a
  complex simple Lie group,
\newblock Amer. J. Math. {\bf 81}, 973--1032 (1959).

\bibitem{bourbaki1968groupes}
N.~Bourbaki,
\newblock Groupes et Algebres de Lie: Elements de Mathematique,
\newblock Hermann, Paris, (1968).

\bibitem{aitken2017}
A.~C. Aitken,
\newblock {\em Determinants and matrices},
\newblock Read Books Ltd, (2017).

\bibitem{katz1997mirror}
S.~Katz, P.~Mayr, and C.~Vafa,
\newblock Mirror symmetry and exact solution of 4D $ N= 2$ gauge theories: I,
\newblock Adv. in Theor. and Math. Phys. {\bf 1}(1), 53--114 (1997).

\bibitem{gebert1996toda}
R.~W. Gebert, S.~Mizogushi, and T.~Inami,
\newblock Toda field theories associated with hyperbolic Kac-Moody
  algebra—Painleve properties and W algebras,
\newblock Int. J. of Mod Phys. A {\bf 11}(31), 5479--5493 (1996).

\bibitem{delfino1996non}
G.~Delfino, G.~Mussardo, and P.~Simonetti,
\newblock Non-integrable quantum field theories as perturbations of certain
  integrable models,
\newblock Nucl. Phys. B {\bf 473}(3), 469--508 (1996).

\end{thebibliography}

\end{document}